\documentclass[onecolumn,authoryear]{els-mrw} 
\usepackage{amsmath,amssymb,amsfonts,amsthm,makeidx,graphicx}
\usepackage[rightcaption]{sidecap}
\usepackage{wrapfig}
\usepackage{txfonts}
\usepackage{helvet}

\begin{document}

\chapter{Hot Subdwarf Stars}\label{chap1}

\author[1]{Ulrich Heber}%

\address[1]{\orgname{Dr. Remeis-Sternwarte \& ECAP}, \orgdiv{University of Erlangen-Nuremberg}, \orgaddress{Sternwartstr. 7, D96049 Bamberg}}

\articletag{}

\maketitle

\begin{glossary}[Glossary]
\term{Hertzsprung-Russell diagram (HRD):} Diagnostic diagram to test stellar evolution theory. Absolute magnitude is plotted versus color, or luminosity vs. effective temperature. Morphological features relate to different stages of stellar evolution:
\begin{itemize}
    \item[] \textbf{Main Sequence (MS):} Stars burning hydrogen in their core, the longest time-span in the evolution of a star.
    
\item[] \textbf{Red giant branch (RGB):} Evolved stars with an inert helium core powered by hydrogen burning in a shell. 

\item[]\textbf{Horizontal branch (HB):} Core helium burning stars.

\item[] \textbf{Hot subdwarfs (SD):} Helium burning, low mass star  

\item[]\textbf{Asymptotic giant branch (AGB):} Giant stars powered by hydrogen and helium burning in two distinct shells, inert C/O core.

\item[] \textbf{White dwarf (WD):} Final state of stellar evolution of low mass stars, stabilized by electron degeneracy. They cool on very long-time scales (age of the Universe).
\end{itemize}

\term{Helium main sequence (HeMS):} Theoretical sequence of helium stars as a function of their mass.

\term{Type Ia supernova (SN\,Ia):} Thermonuclear explosion of a white dwarf.

\term{Zeeman effect:} Splitting of atomic spectral lines in the presence of a magnetic field.

\term{Close-binary stars:} Two gravitationally bound stars orbiting the center of mass, which interact and exchange mass between the components during the evolution of the stars, mostly during giant phases.
\begin{itemize}
    \item[] {\textit Roche geometry}: Describes the gravitational potential of a binary star. Mass transfer between components occurs when a star fills its Roche lobe, and the transfer can either be stable or unstable.
    \item[] Unstable mass transfer may create an envelope engulfing both components, forming a common envelope (CE). The envelope may be ejected, creating a short-period binary, or both stellar components merge.
\end{itemize}

\term{Gravitational wave source:} Short-period SD+WD binaries radiate gravitational waves (GW) detectable by future space-based GW observatories.

\term{Stellar atmosphere:} The outer layer of the star from which the electromagnetic spectrum emerges. Modelling the spectrum allows us to derive its chemical composition and other physical properties.

\term{Stellar pulsations:} (Non-)radial oscillations of stars, an important tool to study the stellar interior.  
\begin{itemize}
\item[] {\textit Pressure mode}: Standing acoustic wave. The restoring force is the pressure gradient.
\item[] {\textit Gravity mode}: Standing wave with buoyancy as restoring force.
\end{itemize}

\term{Galactic stellar populations:} The structure of the Galaxy can be described by three components: the thin disk, the thick disk, and the halo.

\end{glossary}

\begin{glossary}[Nomenclature]
\begin{tabular}{@{}lp{34pc}@{}}
sdB & subdwarf B star\\
sdO & subdwarf O star\\
sdOB & subdwarf OB star\\
GC & globular cluster\\
RV & radial velocity\\
LC & light curve \\
EHB & extreme horizontal branch\\
PCEB & post common envelope binary\\
DD & double detonation\\
BLAP & blue large amplitude pulsator\\
BPS & binary population synthesis\\
LAMOST & Large Sky Area Multi-Object Fiber Spectroscopic Telescope\\
SDSS & Sloan Digital Sky Survey\\
 \textit{TESS} & Transiting Exoplanet Survey Satellite
\end{tabular}
\end{glossary}

\begin{abstract}[Abstract]

Hot subdwarf stars are the stripped cores of red giant stars in transition to the white dwarf sequence. The B-type subdwarfs (sdB) are powered by helium fusion in the core, more evolved ones (sdO) by shell burning. Low mass subdwarfs may evolve through this stage without any support by nuclear fusion. Because the loss of the giants' envelopes is likely caused by mass transfer in binaries, hot subdwarf stars are test beds for close-binary evolution through stable and unstable Roche lobe overflow, common envelope formation and ejection as well as mergers. Many classes of hot subwarfs can be identified according to surface composition, binarity, magnetism, pulsation characteristics and population membership, including members of globular clusters. Observed binaries show a wide spread of orbital periods from 20 minutes to more than 1,000 days with white dwarf or main sequence companions. The closest systems qualify as type Ia supernova progenitors and \textit{LISA} detectable gravitational wave sources. High-precision light curves from \textit{Kepler} and \textit{TESS} combined with radial velocity curves are used to derive masses, while asteroseismology adds information on the internal structure, slow rotation, and synchronization.  
\textit{Gaia}'s parallax measurements now allow us to place the stars in the Hertzsprung-Russell diagram and derive stellar parameters by combining them with multi-band photometry. The stellar radius can be determined to high precision this way. Newton's law can then be used to derive masses if accurate surface gravities are available. Large-scale spectroscopic surveys will provide atmospheric parameters for large samples of stars,
allowing the mass distributions for the diverse subtypes to be established. These are crucial for testing binary synthesis models and constraining poorly known parameters such as the common envelope efficiency as well as the critical threshold mass-ratio for mass transfer stability.  

\end{abstract}

\paragraph{Keywords:} Color-magnitude diagrams of stars, Hertzsprung-Russell diagram, B subdwarf stars, subdwarf O stars, chemically peculiar stars, stellar atmospheres, chemical abundances, stellar evolution, stellar oscillations, close-binary stars, stellar mergers, globular clusters, white dwarfs, Type Ia supernovae,  halo stars

\section*{Key points/Objectives}

\begin{itemize}
\item In the Hertzsprung-Russell diagram, hot subdwarfs represent an intermediate evolutionary phase in transition from the red giant branch to the white dwarf cooling sequence.
\item Hot subdwarfs are mostly powered by helium burning, either in the core or in the shell around the C/O core. 
\item Their highly peculiar abundance patterns with strong enhancements of heavy metals such as lead are probably caused by atomic diffusion.
\item Hot subdwarfs form in close-binary systems via stable or unstable mass transfer, common envelope ejection or mergers.
\item The shortest period systems at orbital periods of few tens of minutes are viable candidates as progenitors of type Ia supernovae, and their gravitational wave radiation is likely detectable to future gravitational wave space antennae.  
\item Pulsating stars as well as eclipsing binaries among the hot subdwarf stars allow masses and radii to be derived.
\item \textit{Gaia} parallaxes can be used to derive the distributions of radii, luminosities, and masses for large samples of hot subdwarfs. They will provide the benchmark to test stellar evolution and binary population synthesis models.  
\end{itemize}

\section{Introduction}\label{chap:intro}

\begin{wrapfigure}{r}{0.43\textwidth}
\centering
\includegraphics[width=.43\textwidth]{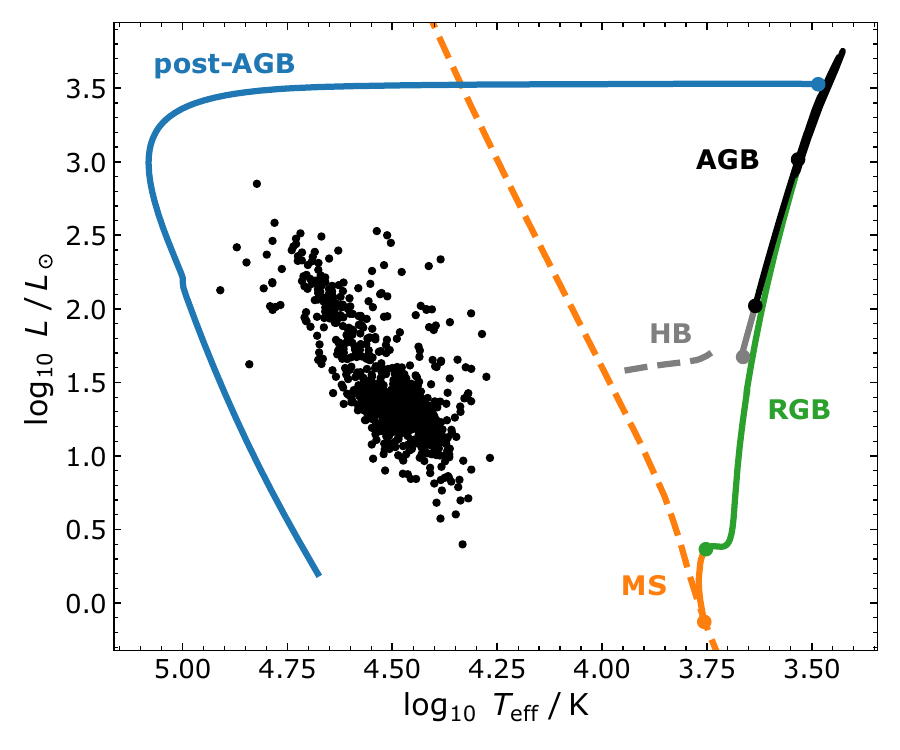}
\caption{Evolution of a Sun-like star in the 
HRD, with evolutionary stages labelled.
Dashed lines represent the zero-age MS
(orange) and the HB (gray). The positions of known SDs within 500pc of the Sun \citep{2024A&A...686A..25D} 
are marked by black dots.
Note that the evolutionary tracks avoid the observed position of the SDs. Courtesy M. Dorsch. 
}
\label{fig:dr2_hrd}
\end{wrapfigure}

\begin{wrapfigure}{r}{0.5\textwidth}
\centering
\vspace*{-1cm}
\includegraphics[width=0.5\textwidth]{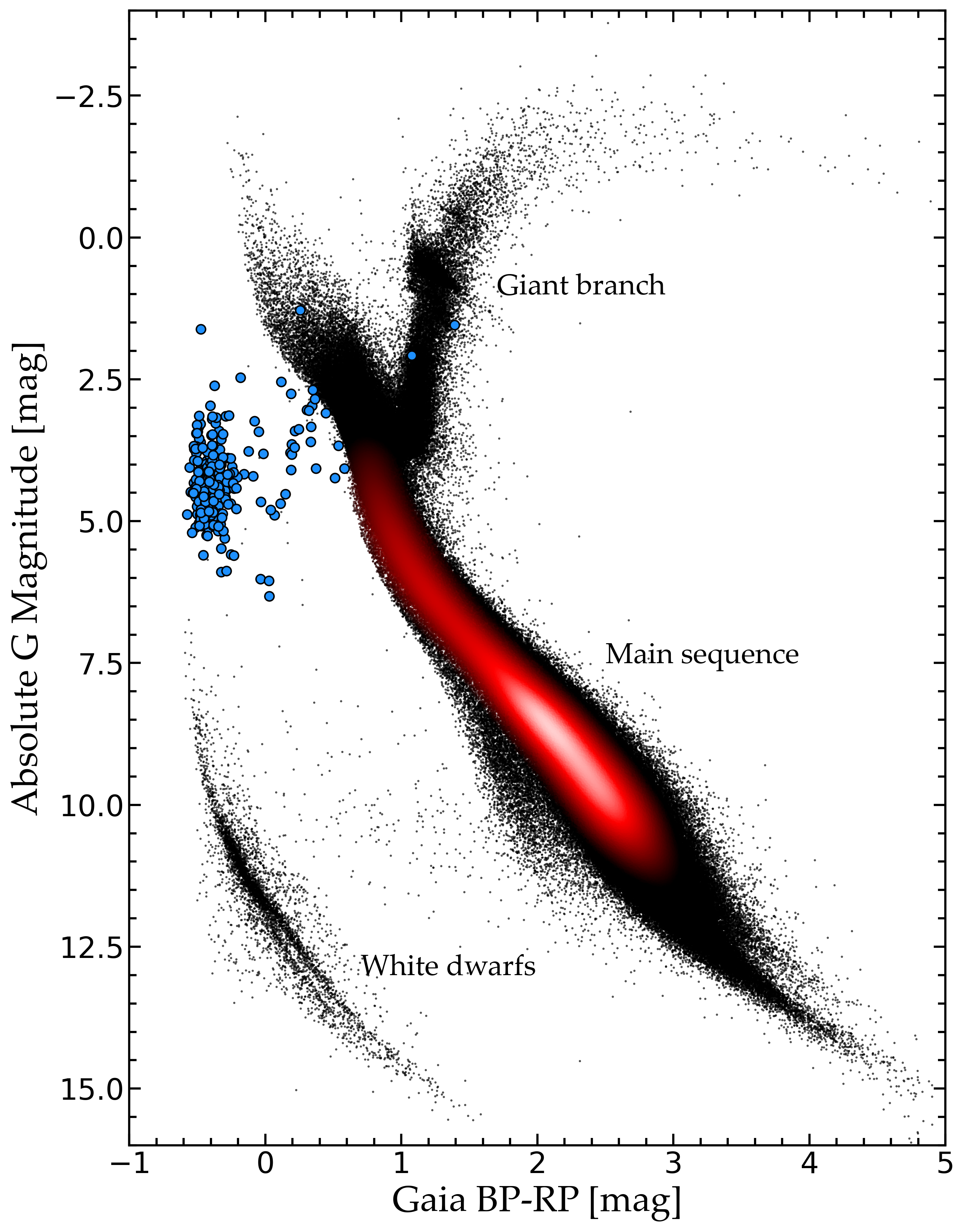}
\caption{Observational HRD derived from Gaia DR3 for stars within 500pc. The blue dots represent SDs from \citet{2024A&A...686A..25D}.
Courtesy: H. Dawson}
\label{fig:hrd_500pc}
\end{wrapfigure}

The evolution of stars is driven by their thermonuclear energy production and depends on the stellar mass. As a star ages, the internal chemical composition changes accordingly. The fundamental diagnostic diagram in stellar astrophysics is the Hertzsprung-Russell diagram (HRD), where luminosity is plotted versus effective temperature (see Fig. \ref{fig:dr2_hrd}) or as absolute visual magnitude versus color (the observer's version, Fig. \ref{fig:hrd_500pc}). Stars start their lives on the main sequence (MS) burning hydrogen to helium in the core.

After core hydrogen exhaustion, the stars expand to become red giants burning hydrogen in a shell (red giant branch, RGB) around the core until helium burning starts, which places the stars on the horizontal branch (HB), mostly a red clump structure superimposed on the RGB. After core helium exhaustion, the star expands again to giant dimensions powered by two shells of energy production, burning hydrogen (outer shell) and helium (inner shell). Energy production by thermonuclear fusion beyond helium occurs only in stars more massive than $\approx$ 8 M$_\odot$. Because hot subdwarf (SD) stars are of low mass, we shall refrain from discussing the evolution of massive stars.

White dwarfs (WDs) are the degenerate final states of stellar evolution of stars born with masses of less than $\approx$ 8 M$_\odot$. Most WDs are the remaining C/O cores of asymptotic giant branch (AGB) stars after ejecting their envelopes near the tip of the AGB. 
However, the transition of giant stars to WDs is not restricted to the AGB, but may happen in earlier stages of giant star evolution. Hot subdwarf stars have been identified as a stellar population linking stars on the RGB to the WD graveyard. They are the helium-burning cores of RGB stars, which lost almost their entire envelope. 

\begin{wrapfigure}{l}{0.45\textwidth}
    \centering
    \includegraphics[width=0.45\textwidth]{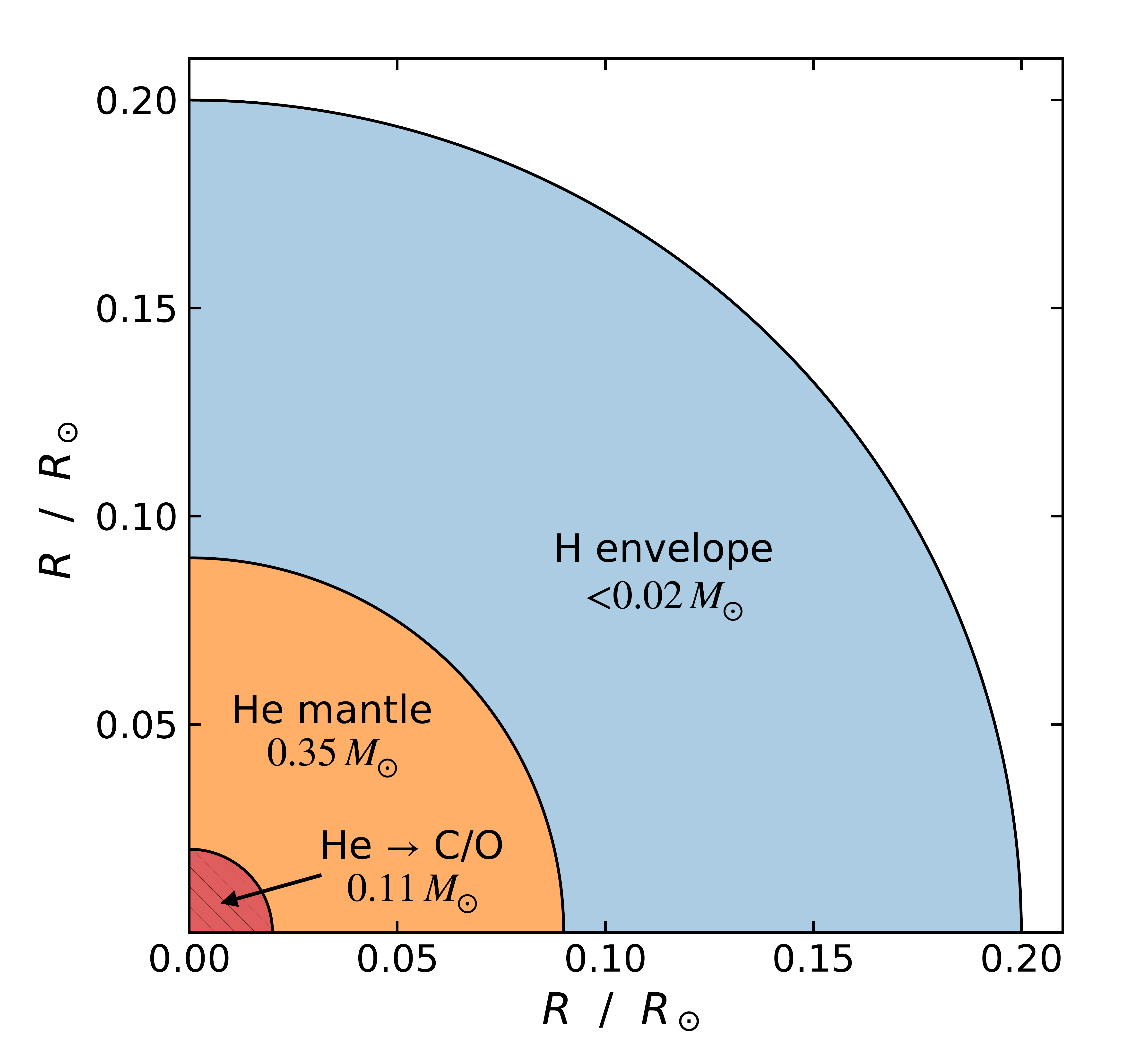}
    \caption{Illustration of the internal structure of an sdB star. The helium burning core (red, hatched) is convective \citep[modified from Fig. 1 of][courtesy M. Dorsch]{2019MNRAS.485.2889P}.}
    \label{fig:sd_structure}
\end{wrapfigure}

Hence, their internal structure (see Fig. \ref{fig:sd_structure}) is similar to HB stars and the term extreme horizontal branch (EHB) stars has been coined to describe their evolutionary state. A major difference between EHB and the much redder HB stars is that their remaining hydrogen-rich envelope is too light ($\lesssim$0.01 M$_\odot$) to sustain the hydrogen-shell burning occurring in normal HB stars. As a consequence, SD stars evolve directly to the WD stage after exhaustion of central helium burning because no nuclear energy production remains to support their luminosity. Most of the known SDs of spectral type B (sdB) can be identified as EHB stars, while many of the hotter O-types (sdO) may have evolved beyond core helium burning and are on their way to become WDs.

Although the SDs are thought to be helium stars, their atmospheric chemical composition spans a wide range of helium abundances. Subdwarf B stars, which are the majority of the known hot SDs, show hydrogen-dominated spectra, with weak to undetectable helium lines, indicative of sub-solar helium abundances spanning several orders of magnitude. On the other hand, the spectra of the majority of the known O-type SDs are helium strong-lined and the contribution of hydrogen to the atmospheric abundance patterns varies enormously from slightly super-solar to being almost devoid.

For AGB stars, envelope-loss can be explained by internal processes involving dust-driven (super-) winds, pulsations, and thermal pulses triggered by the interaction by the H- and He-burning shells.  Because of the enormous radius of an AGB star, its envelope is loosely bound gravitationally to the compact core, and the above-named processes are sufficient to unbind the envelope at the tip of the AGB.
For an RGB star, however, the envelope is hotter, denser and more tightly bound to the core and, therefore, difficult to unbind by internal mechanisms.

An efficient way of envelope removal is stripping by a close companion star via Roche lobe overflow (RLOF). This scenario had been proposed for SD stars in the 1970s \citep{1976ApJ...204..488M},
but surfaced again around the year 2000 when a significant fraction of sdBs turned out to be radial velocity (RV) variable.

Orbital periods range from hours to a few days, indicating the presence of a WD or low mass MS companion.
Because both partners in such systems are separated by a few solar radii, they can only form through shrinkage of the primordial binary orbit, most likely via common envelope (CE) ejection (post common-envelope binary, PCEB).

In addition, about one third of the known SDs are composite spectrum (SB2) binary systems with MS companions of spectral type K to F. Orbital periods are many months to a few years, indicating separations of a few au. Stable RLOF to an MS star at the tip of the RGB has been suggested to form a sdB component in those systems. Another third of SDs does not show any RV variations nor evidence for excess flux from a luminous companion. 
Those SDs are considered truly ''lone wolves''.  In the context of binary evolution, the formation of single stars requires components of progenitor binaries to merge. The most popular scenario is the merger of two low-mass WDs of helium composition, massive enough to ignite helium burning. 
Because the orbital periods of some PCEB systems are very short, their evolution is affected by gravitational wave radiation driving their separation to shrink,
which may finally drive the binary system into a merger. Mass transfer in hot SD binaries with WD companion may trigger the WD to  explode as a thermonuclear supernova (SN Ia) if the WD mass is high enough.
Binary population synthesis (BPS) models have been developed through the last two decades in ever-increasing detail. 
However, BPS modelling is plagued by poorly known physical processes that determine the stability of mass transfer, common envelope formation, evolution and ejection, which might be improved as more observational facts are collected.

To this end, the space missions \textit{Kepler} and \textit{TESS} are strong drivers for the advancement of the field by providing high precision time-series photometry, allowing for detailed 
asteroseismic studies of pulsating SDs as well as 
light curve (LC) analyses of binaries to be carried out. Large ground-based spectroscopic surveys (e.g., SDSS and LAMOST) identified thousands of SDs allowing us to study the SD population as such.  

Finally, ESA's \textit{Gaia} mission has revolutionized stellar as well as Galactic astrophysics. In particular, its astrometric measurements are of unprecedented precision, allowing us to place stars in the 
HRD properly. Its impact on the enhancement of research 
into hot SDs must not be underestimated. Before the second \textit{Gaia} data release (DR2) became available in 2018, trigonometric parallaxes of sufficient precision were available for a handful of SDs, only, measured by the \textit{Hipparcos} mission. For illustration, Fig. \ref{fig:hrd_500pc} shows a HRD for stars within 500pc from the Sun. In the blue part of the HRD, a clump of stars appears at absolute \textit{Gaia} magnitudes M$_\textrm{G}$=3--5 mag, which represent the SD population in this volume. 
The latest Gaia data release (DR3) published astrometric data for more than 1.46 billion stars. 
Among them, there are 6,616 spectroscopically confirmed SDs and more than 60,000 candidates \citep{2022A&A...662A..40C}. 
Although this is a small population of stars compared to that of MS and WD stars, SDs are shorter-lived than the latter at $\approx$ 10$^7$ (post-EHB) to 10$^8$ Myrs (EHB) and, therefore, less common. More importantly, SDs populate a region in the HRD not covered by single star evolution (see Fig. \ref{fig:dr2_hrd}), which represents a yet uncharted region of the HRD and provides an important benchmark for binary star evolution.
Work on SDs before \textit{Gaia} data were released has been reviewed by 
\cite{2009ARA&A..47..211H,2016PASP..128h2001H}.

We start with an overview of atmospheric parameters and chemical compositions of SD stars in Sect. \ref{sect:atmos} and proceed with a discussion of theoretical concepts of and observational evidence for close-binary evolution to explain the formation of SDs in Sect. \ref{sect:binary_evolution}. Single SDs and their potential formation via mergers of He-WDs are discussed in Sect. \ref{sect:merger}. Asteroseismology of pulsating SDs is described in Sect. \ref{sect:pulsation} and stellar parameters, that is radius, luminosity, and mass, are discussed in Sect. \ref{sect:parameters}. We conclude with an outlook (Sect. \ref{sect:conclusion}).

\section{Atmospheric parameters and chemical composition}\label{sect:atmos}

\begin{figure}[t]
\centering
\includegraphics[width=.8\textwidth]{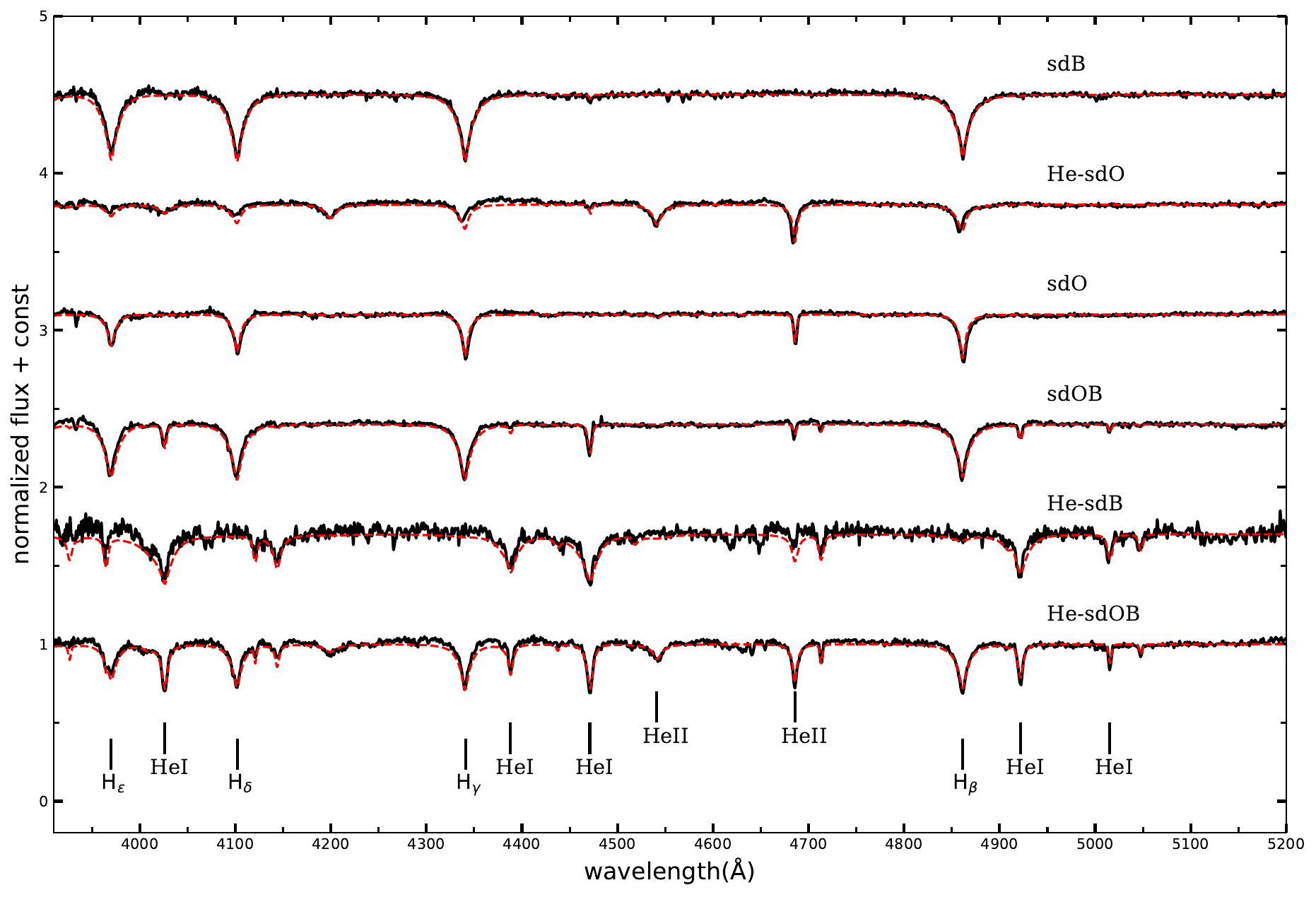}
\caption{Six prototypical normalized spectra (LAMOST DR5, black) of SD classes and best-fitting synthetic spectra (red). Characteristic H and He lines are marked by short vertical lines \citep[modified from Fig. 2 of][courtesy Z. Lei]{2018ApJ...868...70L}.
}
\label{fig:spec_class}
\end{figure}

The optical spectra of sdB and sdO stars cannot be classified in the MK classification system because of their peculiar helium spectra. They can roughly be distinguished into six groups 
\citep[][see Fig. \ref{fig:spec_class}]{2018ApJ...868...70L}.
 The sdB stars are characterized by strong Balmer lines and weak He {\sc{i}} lines, He {\sc{ii}} lines being absent, which, however, are prominent in sdO stars. An intermediate group, sdOB stars, shows weak He {\sc{ii}} 4686$\AA$\ absorption in addition to the He {\sc{i}}. Among the SDs, He line strengths differ enormously. The He-strong ones are, therefore, labelled with the prefix 'He', some of which are seemingly almost devoid of H. It turned out, that HesdOB stars have quite different metal abundance patterns, which distinguish extreme (eHesdOB) from intermediate (iHesdOB) HesdOBs, when their He abundance (by number) exceeds four times their H abundance.

\subsection{Model atmospheres}

Stellar atmospheres are characterized by their effective temperature, surface gravity and chemical composition.
These parameters can be derived from spectroscopic observations using sophisticated model atmospheres. While most normal stars show abundance patterns similar to the Sun, SD stars display many peculiarities. Helium is the second most abundant element in normal stars at an abundance of one tenth of hydrogen. In the atmospheres of some SD stars hydrogen is replaced almost entirely by helium, while in others helium is undetected, that is its abundance can be less than 10$^{-4}$ that of hydrogen by number. Therefore, the He abundance is an important model parameter that impacts the temperature-density stratification of SDs' atmospheres. Likewise, the abundances of metals affect the model properties. Deficiencies of lighter metals and large enrichment of heavy metals are observed. The cumulative effect of the many atomic line transitions arising from different metal species on the atmospheric structure is called metal line blanketing. Because the abundance patterns vary from star to star, usually an averaged composition is adopted. The simplest approach to modelling stellar atmospheres is to assume that locally the layers are in thermodynamic equilibrium (LTE). For early type MS stars, deviations from LTE (NLTE) occur because of the high temperatures involved, in particular for the low densities of the outer layers. In SD stars, the densities are higher, diminishing the departures from LTE. In realistic models both metal line blanketing and NLTE effects need to be dealt with, which is numerically costly, in particular if large numbers of spectral lines need to be included. 
Depending on the parameter regime under study, one, or the other effect may be more important. To be able to deal with metal line blanketing in as much detail as possible, a hybrid approach has been developed that uses an LTE model structure and then computes NLTE departure 
coefficients to synthesize the atmospheric spectrum \citep{2021A&A...650A.102I}. Alternatively, the NLTE model atmosphere code TLUSTY \citep{1995ApJ...439..875H} has often been used for the analysis of SD spectra \citep[e.g.][]{2012MNRAS.427.2180N} with some limitations with respect to metal line blanketing. A detailed discussion can be found in \cite{2011JPhCS.328a2015P}. 
Large-amplitude pulsators (see Sect. \ref{sec:blaps}) produce shocks in their atmospheres during the compression phases, which need to be dealt with when modelling the spectrum \citep{2022MNRAS.515..716J}.

\subsection{Effective temperatures, surface gravities, and helium abundances}

Over the years, quantitative spectral analyses of large samples of nearby SD stars (mostly within 2 kpc from the Sun) have been carried out. \citet{2016PASP..128h2001H} summarized the results for four such surveys, which showed similar parameter distributions in the temperature-gravity and the temperature-He abundance planes \citep[see Fig. 4 in][]{2016PASP..128h2001H}.
Recently, the results of the largest homogeneous sample (more than 1,500 hot SDs) at distances between 500 and 1500\,pc from the Sun have been published \citep{2021ApJS..256...28L} 
based on low-resolution LAMOST spectra and TLUSTY NLTE model atmospheres.
The parameter distributions from LAMOST are fully consistent with previous ones.
\citet{2021ApJS..256...28L} also investigated the kinematics of the stars from \textit{Gaia} DR2 astrometry and were able to assign the stars to the Galactic populations, demonstrating that hot SDs are found in all three populations, which are the Galactic thin disk, thick disk and the halo, although the stars are relatively close to the Sun. 

\begin{figure}[t]
\centering
\includegraphics[width=.49\textwidth]{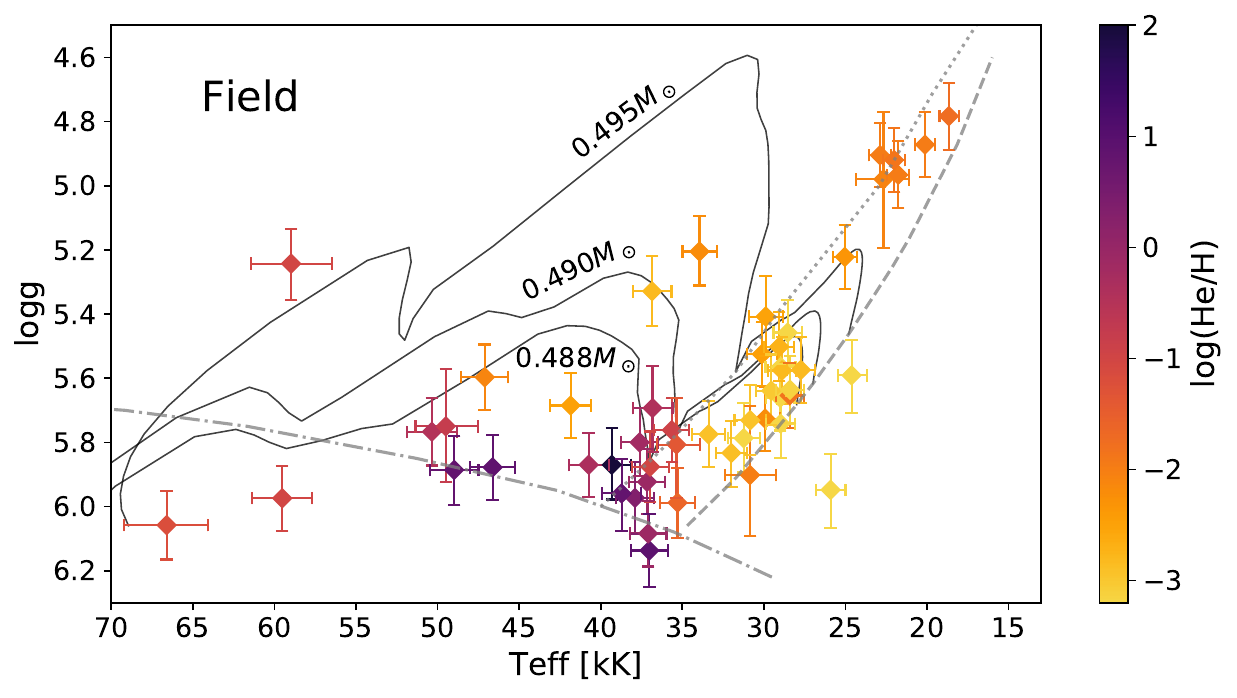}
\includegraphics[width=.49\textwidth]{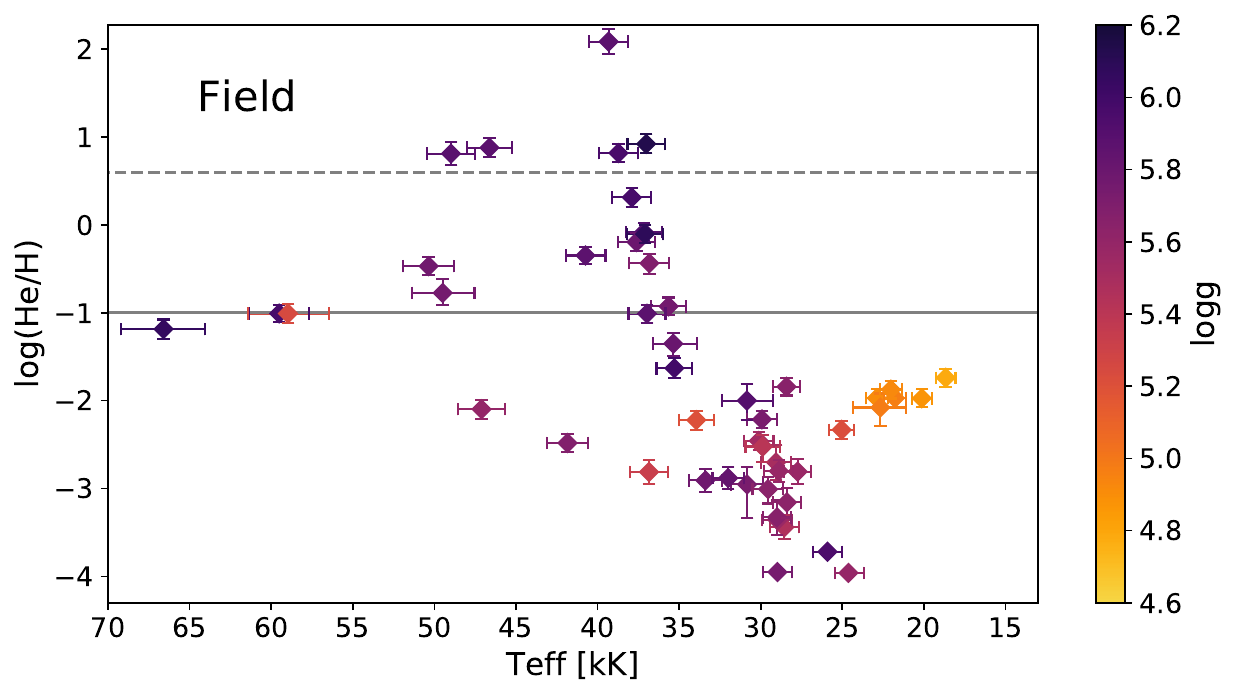}
\includegraphics[width=.49\textwidth]{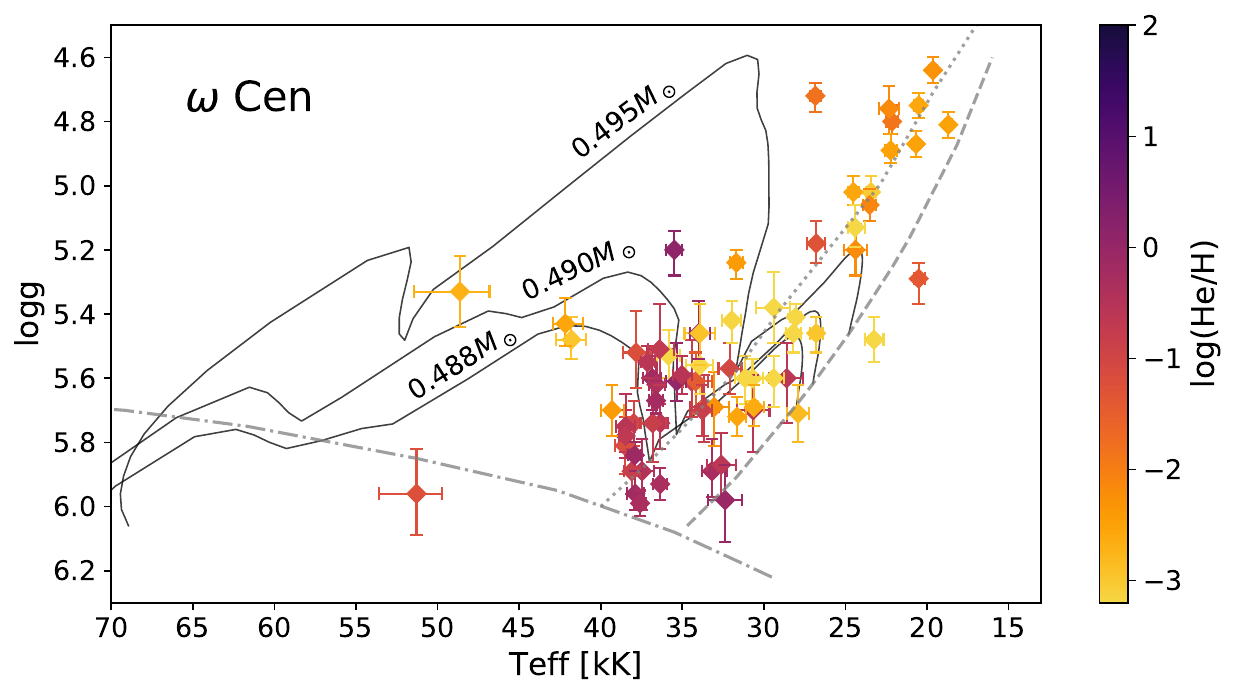}
\includegraphics[width=.49\textwidth]{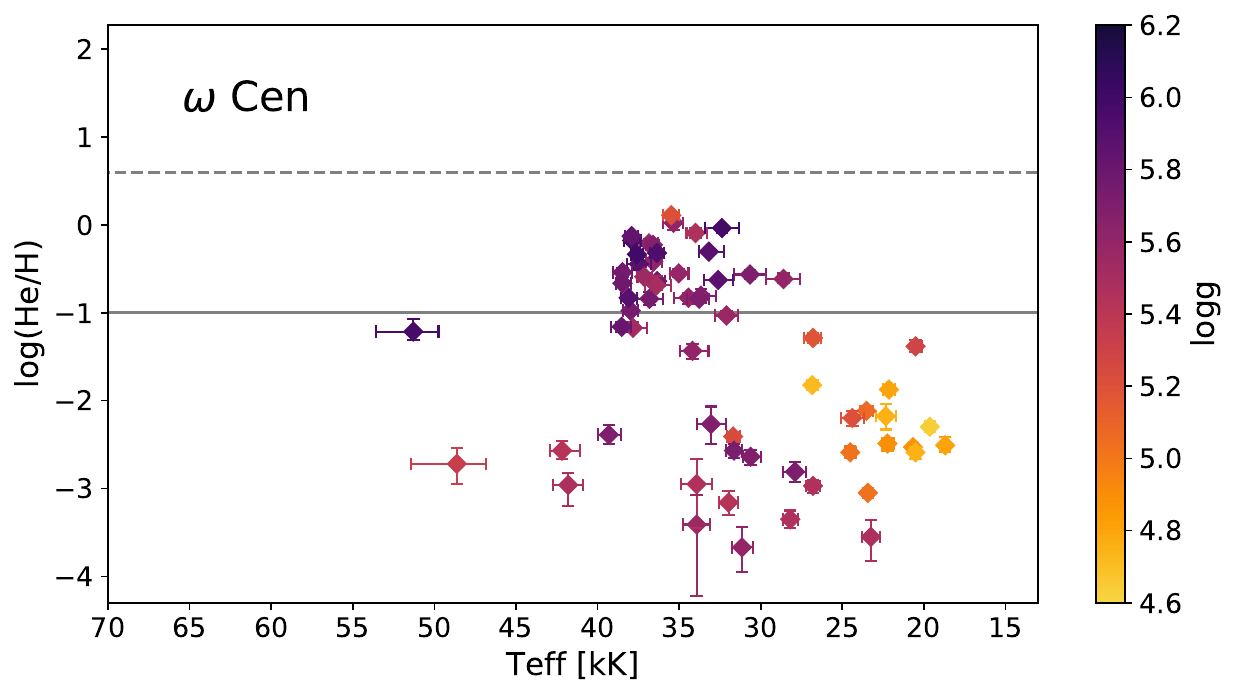}
\caption{Results of spectroscopic analyses of hot SDs in the Galactic halo. Field stars (upper panels) have been studied by \citet{2024arXiv240704479G}; SDs in the globular cluster $\omega$ Cen (lower panels) by \citet{2023A&A...677A..86L}.
{\bf Left-hand panels:} $T_{\rm eff}-\log{g}$ diagrams: The zero-age and terminal-age EHBs for a subsolar metallicity of [Fe/H] = $-1.48$ are marked as dashed and dotted lines, respectively, while the He MS is shown as a dashed-dotted line. The color scales with the He abundance. He-rich objects are marked with filled symbols. Evolutionary tracks for EHB stars with different envelope masses are overlaid as black lines and labeled with the total mass \citep{1993ApJ...419..596D}. 
{\textit{Upper left:}} The sample of halo SDs in the Galactic field \citep{2024arXiv240704479G}.   
{\textit{Lower left:}} Same, but for SDs in $\omega$ Cen \citep{2023A&A...677A..86L}.
{\bf Right-hand panels:} He/H ratio vs. $T_{\rm eff}$.  
Solar helium abundance is marked by the solid horizontal line, while the dotted line marks the transition between intermediate and extreme helium abundance at $\log{n({\rm He})/n({\rm H}})=0.6$. Color scales with the surface gravity. 
{\textit{Upper right:}}
The sample of halo SDs in the Galactic field: 
{\textit{Lower right:}} Same, but for SDs in $\omega$ Cen. There are obvious differences between the distribution of the halo field stars and that in $\omega$ Cen. iHesdOBs are more abundant in the cluster than in the field, whereas there is not a single eHesdOB present in $\omega$ Cen.
Based on data from Table 2 of \citet{2024arXiv240704479G} and 
on Table B2 of \citet{2023A&A...677A..86L}, courtesy L. Kufleitner.  
}
\label{fig:fast_sd}
\end{figure}
\begin{SCfigure}
    \centering
    \includegraphics[width=0.62\linewidth]{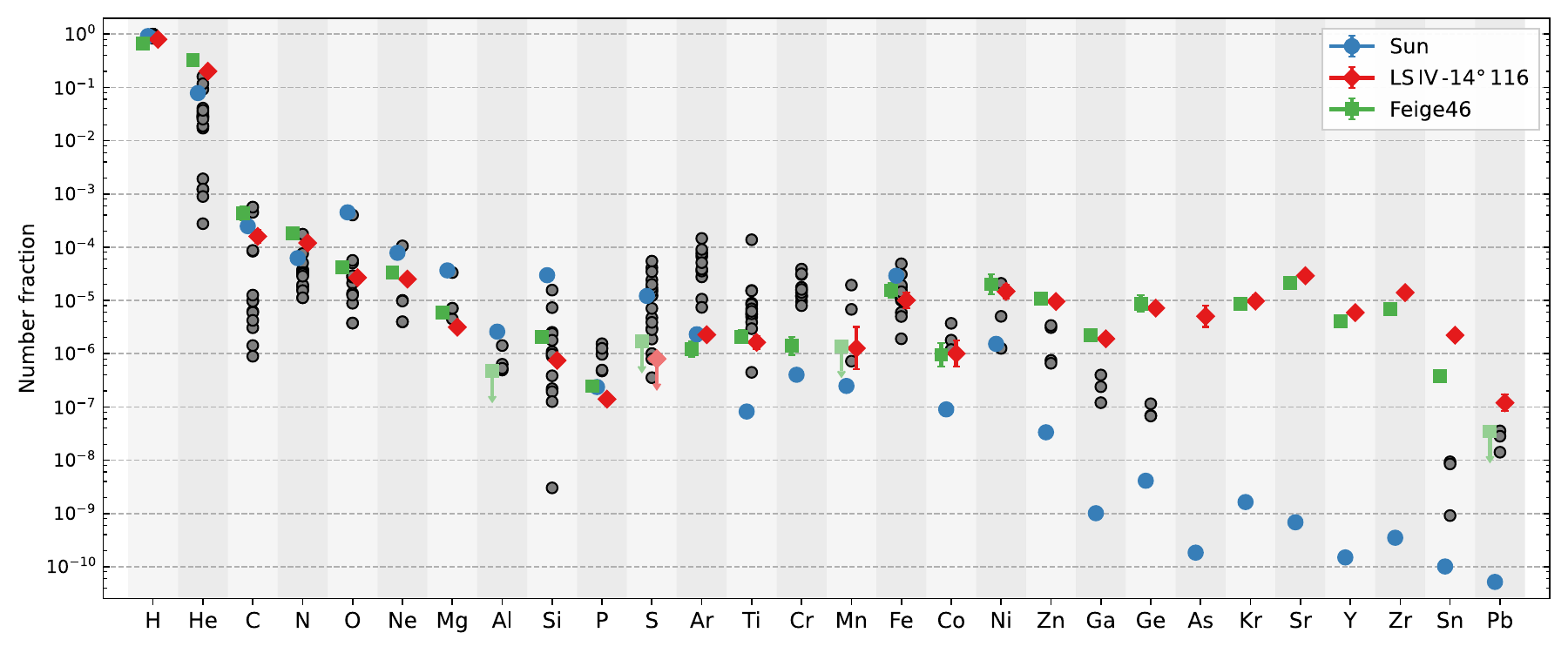}
    \hspace*{0.011\textwidth}
\caption{Abundance patterns (number fractions) of two heavy metal iHesdOBs, LS IV–14$^\circ$116 and Feige\,46 \citep{2020A&A...643A..22D}. 
The abundance patterns of He-poor sdOB stars 
are shown as gray open circles. Note that the abundance patterns of the heavy-metals stars are similar to that of H-poor sdOBs up to and including the iron group elements. While for the Sun the number fractions of heavy metals (Zn, Ga, Ge, As, Kr, Sr,  Y, Zr, Sn, and Pb) level off with increasing atomic number, they stay at a similar level in the heavy-metal SDs \citep[based on Tables 5 and 6 of][courtesy M. Dorsch]{2020A&A...643A..22D}}\label{fig:metals}
\end{SCfigure}

\subsubsection*{Hot subdwarfs in the Galactic halo}

To study more distant  
stars \cite{2024arXiv240704479G} selected the kinematically fastest SDs via \textit{Gaia} proper motions, a sample of faint and distant (2.5 to 15 kpc) halo stars, using SDSS and follow-up spectra obtained with various telescopes. In Fig. \ref{fig:fast_sd} the resulting atmospheric parameters are compared to canonical models \citep{2003MNRAS.341..669H} for EHB evolution.
Again, the distributions for this sample of faint and distant stars are strikingly similar to the ones studied for the 
before-mentioned mixed Galactic population samples.

Globular clusters belong to the halo population. It is a long-standing riddle that globular clusters show diverse horizontal branch morphologies. To characterize HB morphology, the RR Lyr regime of the HB is used as reference. Most clusters' HBs are dominated by stars to the red of the RR Lyrs (RHB), while only few have pronounced BHBs, which are bluer than the RR\,Lyrs, of which a few extend to the EHB (e.g., NGC\,6752 and $\omega$ Cen), that is they host SD stars. The SD population in NGC\,6752 differs considerably from that of the field stars because no He-rich SD has been found, whereas a population of He-rich SDs exists in $\omega$ Cen forming a clump of stars at the hot end of the EHB  \citep{2023A&A...677A..86L} similar to the halo field population
(see Fig. \ref{fig:fast_sd}). While the population of SDs along the EHB band is similar in both populations (field and globular cluster $\omega$ Cen), sdOs are rare in $\omega$ Cen when compared to the field population of halo SDs.
In addition, extreme HesdOBs are missing in $\omega$ Cen. This adds to the diversity of HB properties in GCs, yet to be explained.

\subsection{The diversity of metal abundance patterns}\label{sect:metals}

{
The determination of metal abundances from optical spectra of hot SDs is limited to few species of low atomic mass (e.g. C, N, and Si) because of the lack of spectral lines from other elements. Therefore, UV spectroscopy is required to derive more detailed abundance patterns (see Fig. 6). Fortunately, a plethora of lines from heavy metals, that is the iron group elements and beyond, can be studied from high-resolution UV spectra, which are not accessible from optical spectra. UV spectral analyses are scarce and, therefore, detailed knowledge of abundance patterns is incomplete. However, many abundance analyses of optical spectra are available for most of the subtypes \citep[e.g.][]{2024ApJS..271...21L}. The metal abundance patterns of hot SDs are diverse, with large star-to-star spreads.}
In the He-poor sdB and sdOB stars, the helium deficiency is accompanied by subsolar abundances of elements up to sulfur. The elements of the iron group, with the notable exception of iron, are enriched by 10 to 100 and so are the trans-iron elements (see Fig. \ref{fig:metals}). The abundance pattern of the iHesdOBs is even more exotic. The pattern is very similar to that of the He-poor ones for the light metals and the iron group, but the enrichment of the heavy elements such as Y, Zr, Sn, and Pb (Fig. \ref{fig:metals}) is more dramatic, that is 10$^4$ to 10$^5$ times solar. 
The origin of these trends is usually attributed to diffusion, that is, the interplay of gravitational settling and radiative levitation via the plethora of UV spectral lines. Trans-iron elements may also result from neutron-capture nucleosynthesis. However, nuclear reactions that produce neutrons are known to occur during the thermally pulsing (TP) phase of AGB evolution, but are not expected to operate in earlier stages of evolution. \cite{2023A&A...680L..13B} explored the possibility that neutron production by convective mixing of H into He burning regions, similar to the nuclear reactions in TP AGB stars, could occur during the He-flash to provide neutrons via proton caption by $^{12}$C and subsequent $^{13}$C($\alpha$,n)$^{16}$O, which might lead to heavy-element enrichment as large as observed in the iHesdOBs.  

The eHesdOs seem to be exceptional because they
can be divided into N-rich, C-, and C\&N-rich subgroups \citep{2016PASP..128h2001H,2024ApJS..271...21L}, where the N-rich stars are found among the cooler ones, near $\approx$ 40,000\,K, while the C-rich are hotter (see Fig. \ref{fig:sdO_distribution}). The N enrichment may be caused by nucleosynthesis via the CNO-cycle, and the C enrichment from additional 3$\alpha$ He fusion. Such compositions could be produced in He WD mergers (see Sect. \ref{sect:merger}).

\section{Hot subdwarfs as test beds for binary star evolution}\label{sect:binary_evolution}

While binary evolution scenarios for the formation of SDs have been developed early on \citep{1976ApJ...204..488M,1984ApJ...277..355W}, observational evidence for a large fraction of close binaries among SDs emerged only by the turn of the century \citep{1998ApJ...502..394S,2001MNRAS.326.1391M}, see Sect. \ref{sect:binaries_observations}, which in turn led to a revival of binary evolution concept studies.

Binaries form with different mass ratios and separations, some at more than 10$^4$ au. For such large separations, each of the stars may evolve more or less independently. More interesting are close-binary systems, in which interaction between the components occurs during the evolution of the components. 
In such systems, tidal forces can deform the shape of the stars, affect their rotation, as well as the binary orbit. The more massive primary star leaves the MS first by expanding to giant dimensions and may fill its Roche lobe. Material is then transferred to the less massive secondary through the inner Lagrangian point. The evolution of the system depends on the stability of the mass transfer. If the mass transfer is dynamically stable, most of the primary's envelope is transferred to the secondary via Roche lobe overflow (RLOF), while the secondary becomes more massive because of the accreted mass, which depends on whether the mass transfer is conservative or not. If the mass transfer is dynamically unstable, the mass transfer rate increases so much that the secondary cannot accrete all the transferred material, which piles up until it overfills the secondary’s Roche lobe and engulfs both components to form a CE. 
The secondary is dragged towards the primary's core
and the orbital period decreases. Because of the friction with the CE, orbital energy of the engulfed binary is transferred to the CE. If this energy is large enough to unbind the CE a SD, that is, the core of the former primary, will emerge orbiting an MS star with a short period of hours to few days. However, if the CE cannot be ejected, the binary components will merge into a single star.

\subsection{Formation channels: Stable RLOF versus CE evolution}

\begin{SCfigure}
\centering
\includegraphics[width=0.75\textwidth]{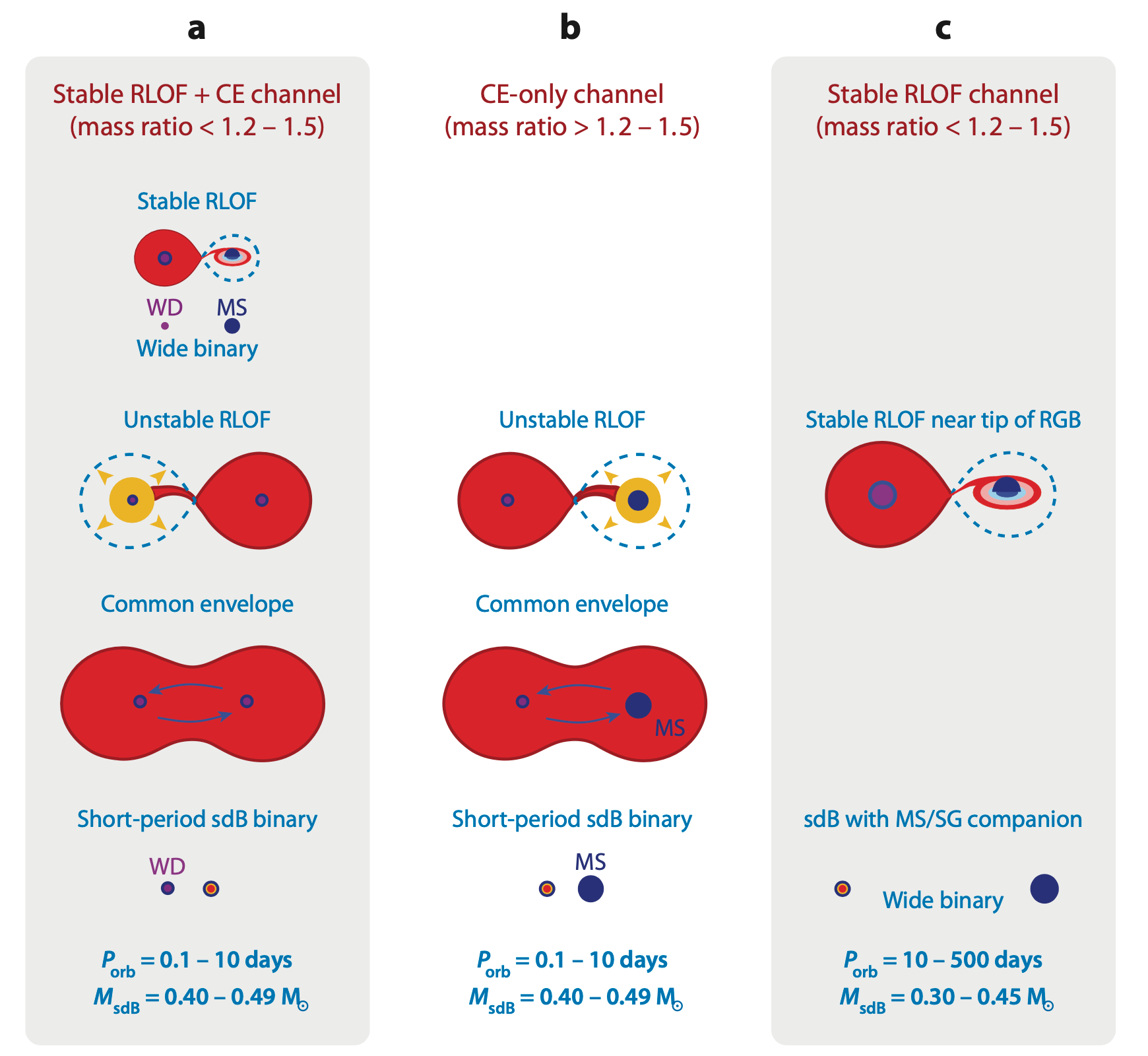}
\hspace*{0.011\textwidth} 
\caption{Three schematic formation channels of sdB stars in close binaries \citep{2008ASPC..392...15P}. 
 (a) For small initial mass ratios, that is $q=M_1/M_2<$1.2 to 1.5, the first RLOF is stable, while  the second one is unstable and a CE is ejected resulting in an sdB+WD with a short orbital period ($\approx$0.1 to 10 d). (b) For $q>$1.2–1.5 already in the first mass-transfer phase the CE is ejected, producing an sdB+MS system in a short-period orbit ($\approx$0.1 to 10d). (c) For low initial $q$ the first RLOF is stable and a sdB forms in a wide, long-period orbit  ($\approx$10 to 500 d or more) with an MS star or subgiant companion \citep[reproduced from][copyright IOP] {2016PASP..128h2001H}.}
\label{fig:cee_rlof}
\end{SCfigure}

CE ejection is often modeled with an energy formalism to constrain its outcome \citep[see][for a review]{2024PrPNP.13404083C}. The binding energy is calculated by tracing the change of the donor's total energy as a function of the remnant mass. However, several physical model parameters, such as the CE efficiency $\alpha_\textrm{CE}$, are poorly known. Short-orbital-period sdB+WD systems likely form after two mass exchange phases, a first stable one and a second unstable CE phase (see Fig. \ref{fig:cee_rlof}). After the first mass-transfer phase, the well-known orbital period -- WD mass (P$_\textrm{orb}$ -- M$_\textrm{WD}$) relation holds, which results from the mass-radius relation for giants with electron-degenerate cores. The latter depends strongly on metallicity.
Whether mass transfer is stable or not depends on the mass ratio $q=M_1/M_2$ at the onset of RLOF
{
($M_1$ and $M_2$ are the masses of the primary, that is the initially more massive, and the secondary stars, respectively.).}.
 For low mass ratios, the mass transfer is stable. However, the threshold q$_\textrm{crit}$ is uncertain and usually is adopted to be between 1.2 and 1.5, independent of the progenitor mass of the sdB. To constrain this parameter is a challenging task, though, because the progenitor's mass, initial and final mass ratios, and orbital periods before and after the CE have to be known as well as the remnant's mass. 
An attempt to derive constraints on $\alpha_\textrm{CE}$ from $\approx$ 150 SD binaries has been published by \citet{2022ApJ...933..137G} 
and extended by including additional model parameters from seven SD+WD binaries with known inclinations \citep{2024ApJ...961..202G}. 
In addition, radii of evolved red giants, and hence the resulting orbital periods, strongly depend on their metallicity.

\subsection{Observational evidence for binarity}\label{sect:binaries_observations}

Evidence for a large fraction of SD binaries has steadily increased since the turn of the century \citep[see][for an overview]{2016PASP..128h2001H}. 
Two groups of SDs are known to be in binaries. In about half of the single-lined SDs, RV variations with periods of a few tens of minutes to a few days have been detected. Composite spectrum SDs do not show such short-period RV variation, but their spectra display absorption lines of neutral or low-ionization species (e.g., Mg {\sc i}, Ca {\sc ii} triplets) arising from F-, G- or K-type MS companions. Whether such systems formed via interaction can be revealed from measurements of the components' separations.

\subsection{The scarcity of common proper-motion pairs and hierarchical triple systems}\label{sect:cm_triples}

The absence of short-period RV variations of composite spectrum SDs could imply that the components could be spatially resolvable.  A first attempt \citep{2002A&A...383..938H} based on \textit{HST}  
observations of 19 systems succeeded to resolving two, only, although about one third of the observed sample should have been, if the separation distribution of progenitors in binary systems were normal. This evidence for scarcity of wide systems has been further strengthened by a recent astrometric study.
\textit{Gaia's} superb spatial resolution allows binaries with separations $>$20 mas to be resolved, which corresponds to orbital periods of at least some tens of years for typical distances of known SD+FGK-MS systems of $\approx$1 kpc. Because orbital movement of the components is small, stars in such systems should be co-moving in space, that is, they should be detectable as common proper-motion pairs. However, the fraction of SD binaries with resolved co-moving stars \cite[$\approx$0.05\%,][]{2020A&A...642A.180P}    
is small, much lower than that of the progenitors. 
By closer inspection, 6 out of the 19 common-proper motion binaries studied by \cite{2020A&A...642A.180P} turned out to be hierarchical triples with an inner close SD binary.
\cite{2022MNRAS.517.2111P}  
using population synthesis techniques followed the stellar, binary, and gravitational dynamical evolution of triple-star systems and found that, indeed, SD stars in binaries with wide, non-interacting companions can be formed. The wide range of possible interactions allow binary and even single SDs to form by triple-star evolution through mass-transfer and merger phases. The large parameter range for triple configurations still has to be explored.   

\newpage
\subsection{Unresolved composite spectrum SDs}

\begin{wrapfigure}{R}{0.5\textwidth}
\centering
\includegraphics[width=0.5\textwidth]{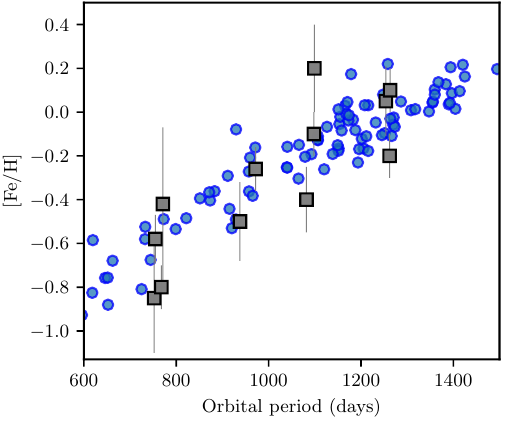}
\caption{Orbital period versus metallicity ([Fe/H]) for long-period sdB composite spectrum binaries.
Observed results are also compared to predicted ones. The results for the observed sample are shown as black squares, whereas those from various BPS models are shown by blue circles. The correlation between [Fe/H] and the orbital period is very well reproduced by the BPS models \citep[see][for details]{2020A&A...641A.163V}. 
Reproduced with permission by A\&A from \citet[Fig. 9]{2020A&A...641A.163V}; copyright ESO.
}
\label{fig:vos_composites}
\end{wrapfigure}

Complete spectral analyses from high-resolution spectra are available for about two dozen systems, including the determination of orbital parameters \citep{2018MNRAS.473..693V}.
If the SD+FGK-MS systems cannot be resolved, their orbits could be derived by
long-term RV monitoring. In a painstaking effort, \cite{2018MNRAS.473..693V} were able to measure periodic small-amplitude RV variations on time scales of a few years, indicating orbital separations of a few au, indeed too small to be resolved, but sufficiently close for stable RLOF mass transfer at the tip of the RGB. 
The metallicities of the cool companions turned out to be low, ranging from [Fe/H]=-0.8 to 0.0 (solar), decreasing with decreasing orbital period. Hence, the systems belong to old stellar populations. Two bona fide halo members with low mass companions were found recently, at metallicities of -1.04 and -1.9 \citep{2021A&A...653A...3N,2021A&A...653A.120D}. 

In addition, a correlation of metallicity with orbital period was discovered, as well as one with the mass ratio \citep{2020A&A...641A.163V}. 
Their detailed BPS study included the Galactic chemical evolution to address the range of observed low metallicities.
The observed period-mass ratio and period-metallicity correlations were matched by synthesized stellar population predictions very well (see Fig. \ref{fig:vos_composites}). 
Tidal forces tend to circularize the orbit. Therefore, the orbits of these binaries should be circularized before the onset of RLOF according to predictions of canonical stellar evolution theory. 
However, orbits were found to be eccentric and the eccentricity $\epsilon$ to increase up to $\epsilon$=0.25 with increasing orbital period. The eccentricities could be reintroduced during the mass transfer phase \citep{2015A&A...579A..49V}. Such eccentricity pumping was modelled in phase-dependent RLOF assuming that a circumbinary disk has piled up. The models, however, predict eccentricities to decrease with orbital period, opposite to the observed trend.

\subsection{Single-lined SD binaries}

The literature on single-lined SD binaries is extensive \citep[see e.g.][]{2016PASP..128h2001H, 2012ApJ...753..101B,2022A&A...666A.182S,2023A&A...673A..90S,2024A&A...686A.126B}. RV variability of apparently single SDs with periods between about an hour to several days is common \citep{2022A&A...661A.113G}. 
In short, unseen companions to SDs can be best studied from combining RV and LC analyses. RV monitoring of apparently single SDs revealed periodic variations with semi-amplitudes as high as $\approx$ 500 km\,s$^{-1}$.
As indicated by their mass functions, the unseen companions are either low-mass (M-type) MS stars, or brown dwarfs, or WDs.
High precision LCs come from the \textit{Kepler} mission (almost uninterrupted photometry with cadences of 29.4 or 1 min over a 4-yr time span or a few months, respectively). The Transiting Exoplanet Survey Satellite (\textit{TESS}) mission surveys the sky in a series of 27-days long observing sectors with cadences of 2 min or 20s, respectively, and is going on to measure LCs of almost half of the known SDs to unprecedented precision. 
Their analyses consider eclipses, reflection effect, and ellipsoidal variations to determine the stellar parameters. The high photometric precision allows tiny effects such as Doppler boosting, gravitational lensing, as well as the R\o mer delay to be observed and studied. Some important recent discoveries shall be discussed.

\subsection{Roche-lobe filling SD binaries, massive companions, and circumbinary planets}

As time goes by, the system evolves and the orbit shrinks due to gravitational wave emission. Eventually, one of the components will fill its Roche lobe to start mass transfer, which will be the SD if the companion is a WD. Indeed, 
two ultra-compact SD binaries in very short-period orbits (39 and 56 min, respectively) with WD companions were discovered recently \citep{2020ApJ...898L..25K}, in which the SDs fill their Roche lobes and transfer mass at a rate of $\approx$10$^{-9}$ M$_\odot$\,yr$^{-1}$. Direct evidence for an expected accretion disk, though, is elusive \citep{2023MNRAS.519..148D}. 

Massive companions (neutron stars or black holes) in short-period orbits with SDs have long been thought of, but claimed detections are doubtful. A new discovery of a \textit{Gaia} astrometric binary (BPS BS 16981-0016) consisting of a 1.05--1.9 M$_\odot$ compact object in a 892 d orbit around a sdB \citep{2023A&A...677A..11G} comes unexpectedly because of the long orbital period. Nevertheless, evolutionary models \citep{2018A&A...618A..14W} predict such systems to exist; yet another pathway of close-binary evolution has been found.
Up to now, \textit{Gaia} data allowed the detection of a few astrometric binaries, only. Hence, \textit{Gaia} DR3 provided a glimpse of the scientific results that will emerge as Gaia data improve \citep{2024NewAR..9801694E}. 

Very low mass third bodies have also been found around SD binaries.
Long-term photometric monitoring of eclipsing, short-period SD+MS (HW Vir) systems revealed variations of eclipse times, which were attributed to the presence of a third body of planetary mass orbiting the binary. However, doubts about the existence of those planets have grown and other mechanisms such as the Applegate effect have been invoked \citep[see][for a review]{2022MNRAS.514.5725P}. 
On the other hand, circumbinary planets may exist that form in CE evolution as fall-back disks and, thus, are second generation planets  
\citep[see][for a review]{2023A&A...675A.184L}.

\subsection{Supernova type Ia progenitors, high velocity SDs, and gravitational wave sources}

A type Ia supernova (SN Ia) results from the thermonuclear explosion of a WD when its mass exceeds a critical limit, which is often considered to be the Chandrasekhar mass.
The double degenerate merger scenario explains the formation of such a high mass object by a merger of two C/O-WDs of sufficiently high masses. 
An alternative scenario considers a WD+He star binary, in which He is accreted by a massive WD. Once a critical mass of He has been accreted on top of the WD, an explosion of the He shell occurs which
then triggers the explosion of the WD's C/O core (Double detonation, DD). 
In consequence, the mass of the exploding WD may be lower than the Chandrasekhar limit and the resulting SN Ia fainter \citep{2007A&A...476.1133F}.

Close SD+C/O-WD binaries are considered viable progenitors for DD SN Ia because the SD may  provide He to be accreted by the WD once mass transfer started. However, candidate systems are rare, because of the short SD lifetimes ($\approx$100Myr). Because of its very short orbital  period of 
20.5 min, ZTF J0526+5934 \citep{2023ApJ...959..114K} is of particular interest, although its position in the HRD below the EHB indicates that it is of very low mass \citep[0.32--0.36 M$_\odot$,][]{2024NatAs...8..491L} or not He-burning at all
\citep[extremely low mass WD,][] {2024A&A...686A.221R}.
Although ZTF J0526+5934 will merge within 1.4 Myrs, it is not expected to form an SN Ia \citep{2023ApJ...959..114K}.
Only CD-30$^\circ$11223 \citep[P= 70.53 min,][]{2024MNRAS.527.2072D} and
PTF1~J2238+7430 \citep[P=76 min,][]{2022ApJ...925L..12K} are Chandrasekhar mass systems with merging times shorter than the EHB lifetime and, therefore, may explode as DD SN Ia. {
\citet{2024arXiv240517896P} suggest that rotation-induced mixing
of accreting WD may prevent SN Ia in these systems.}

As a result of the explosion of the WD, the binary may be disrupted and the surviving SD donor will be released at a space velocity, occasionally large enough for the star to escape from the Galaxy. Such a hyper-velocity star (HVS), US\,708, was discovered in 2005 \citep{2005A&A...444L..61H}. Simulations suggested that HVS candidates among SDs have been overlooked  
\citep{2022A&A...663A..91N}. Despite an extensive survey, the search for additional unbound SDs was not successful \citep{2024arXiv240704479G}. 

Compact binaries are important sources of gravitational waves, which will be detected by future 
space-based gravitational wave (GW) observatories such as ESA's \textit{LISA} and China's \textit{TianQin}. The GW amplitude depends on mass, distance and orbital period. Because the GW amplitude increases with decreasing period, the shortest period systems are the best candidates to be detected. Today, out of six of the shortest period SD+WD binaries (periods below 100 min), two (CD-30$^\circ$11223 and HD 265435), along with the verification source ZTF J0526+5934 are likely to be detected by both observatories \citep{2024ApJ...963..100K,2024NatAs...8..491L}.

\subsection{Related stripped low mass stars in close binaries}\label{sect:elm}

Stripping of the RGB envelope in a close binary may occur well before the tip of the RGB and the remnant might not ignite helium burning but evolves into a He WD of low mass. If stripping occurs early on in the RGB evolution, when the helium core is still small, extremely low mass (ELM) WDs should form with masses less than 0.3 M$_\odot$ (ELM-WD), which, indeed, have been discovered \citep[see ][for an extensive discussion]{2016PASP..128h2001H}. ELM WDs and their progenitors, thermally bloated WDs (pre-ELM WDs), are found in binaries with WDs \citep{2023ApJ...950..141K}, neutron stars \citep[i.e. pulsars,][]{1996ApJ...467L..89V}, and MS stars of spectral types from late B to F \citep[EL CVn systems,][]{2011MNRAS.418.1156M,2024NewA..10702153P}, which have orbital periods from about a day to several tens of days. A system in an unusually long period (450d) orbit hosting a 0.2 M$_\odot$ ELM WD and a solar-like star was detected via gravitational self-lensing \citep{2019ApJ...881L...3M} challenging binary evolution models.
These systems are analogs to close SD binaries, with the reservation that neutron star companions to hot SDs have yet to be confirmed. Recently, a pre-ELM WD has been discovered in a 3.4h orbit with a brown companion \citep{2021A&A...650A.102I}, which would be an analog to the SD+M/BD systems. 
Multi-periodic oscillations have also been observed in ELM WDs \citep{2016ApJ...822L..27G} as well as in pre-ELM WDs \citep[EL CVn][]{2013Natur.498..463M}.
Most EL CVn appear to be inner binaries of hierarchical triples \citep{2020MNRAS.499L.121L}, a subject still under study
for similar SD systems (see Sect. \ref{sect:cm_triples}).

\section{Single SDs and the role of WD mergers}\label{sect:merger}

Although several scenarios have been proposed \citep[see][]{2016PASP..128h2001H}, the formation of a single SD is far from clear. Recently, the merger concept entered the focus of interest to explain the origin of single SDs, at least that of the He-rich ones. Accordingly, a single star could form via a merger of two He-WDs as first suggested by \cite{1984ApJ...277..355W}. Loss of energy and angular momentum by the inevitable emission of gravitational waves and breaking by magnetic fields, if present, will lead to orbital shrinkage.
Because of the WDs mass-radius relation, the less massive WD is the larger one and fills the Roche lobe first. Because this is a run-away process, the companion is destroyed within a few {
 revolutions.} 

\begin{SCfigure}
\centering
\includegraphics[width=.62\textwidth]{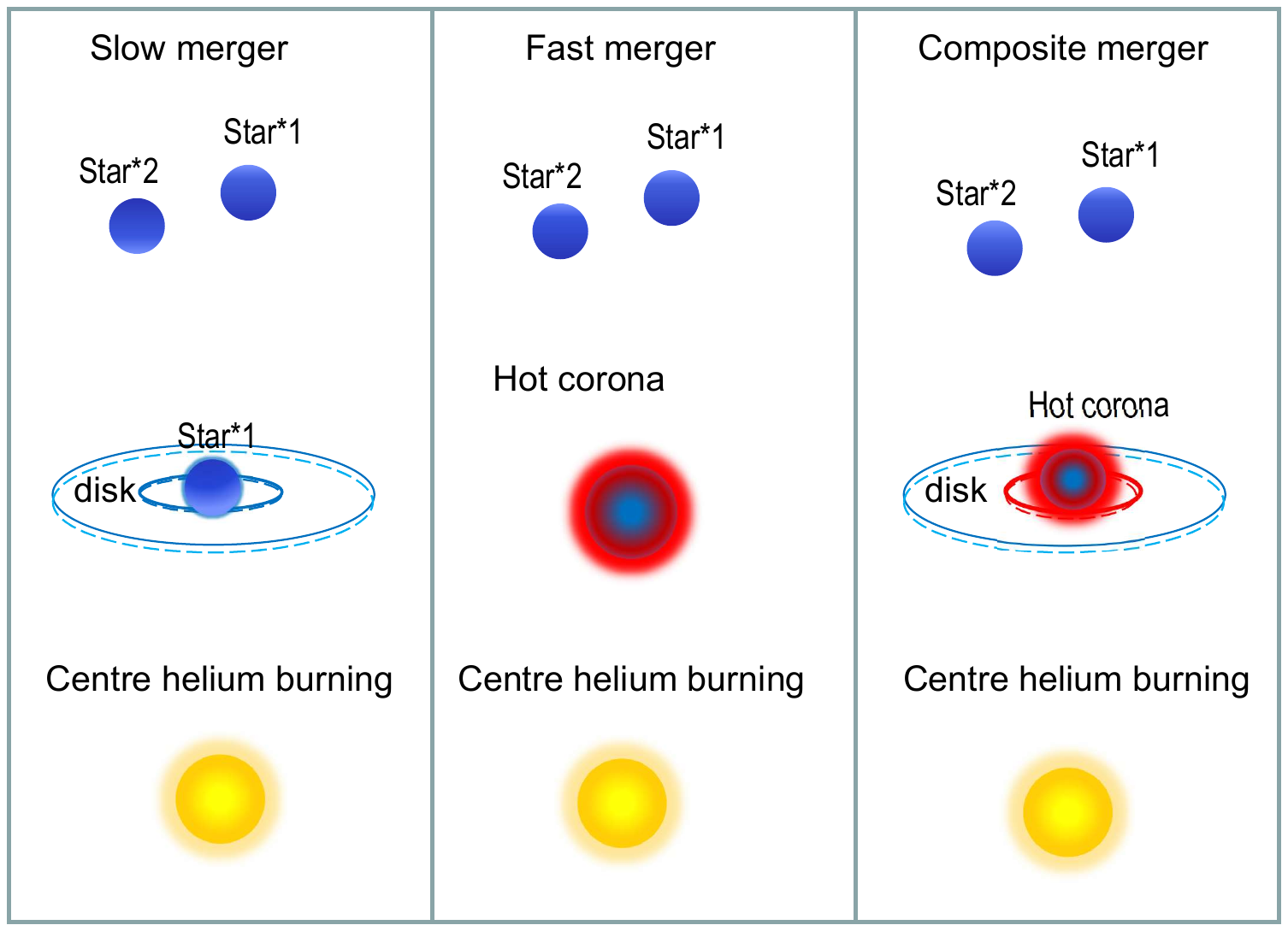}
\hspace*{0.011\textwidth}
\caption{Schematic of three possible ways in which two helium WDs might merge \citep{2012MNRAS.419..452Z}. {\bf Left:} For a low accretion rate 
the merger is slow, forms a disk and temperatures are too low for He burning.
Therefore, the composition of the accreted material from the disrupted He WD is unchanged and remains C-poor and N-rich. 
{\bf Middle:} A fast merger (high accretion rate) lasting a few minutes forms a corona, hot enough to produce C. 
{\bf Right:} In a composite model the merger starts with a fast accretion phase and then proceeds as a slow process to form both a disk and a hot corona.
\citep[modified from Fig. 1 of][courtesy X. Zhang]{2012MNRAS.419..452Z}}. 
\label{fig:merger}
\end{SCfigure}

{
Models for He-WD binary mergers \citep{2012MNRAS.419..452Z} consider three different scenarios (see Fig. \ref{fig:merger}). The properties of the merger product depend on the details of the mass transfer. For a low accretion rate (M$\dot{M}$=10$^{-5}$ M$_\odot$\,yr$^{-1}$) the merger is slow, a disk forms but does not heat up to temperatures high enough for nucleosynthesis to start. Hence, the surface composition of the resulting SD should be the same as that of He-WDs, that is C-poor and N-rich. At a high accretion rate ($\dot{M}$=10$^{4}$ M$_\odot$\,yr$^{-1}$) the merger is fast enough to finish after a few minutes and produces a corona, hot enough (10$^8$K) to produce C via the 3$\alpha$ process. A composite model consists of a fast accretion phase lasting until 0.1 M$_\odot$ has been accreted and then proceeds as a slow process, leading to both a disk and a hot corona. Accordingly, merger models predict that He-WD mergers form either N-rich (slow merger) or C-rich (fast merger) or C\&N-rich (composite merger) He-rich SDs, respectively, if a disk or a corona or both form (see Fig. \ref{fig:merger}). Indeed, two groups of He-rich SDs are observed with C-rich and N-rich chemical composition, the C-rich being hotter than the N-rich (see Sect. \ref{sect:metals} and Fig. \ref{fig:sdO_distribution}).} 
Recent models \citep{2021MNRAS.504.2670Y} predict that N-rich ones should be about twice as abundant as C-rich ones, which needs to be tested from complete samples of SD stars.  
Mergers of He-WDs with MS stars during the CE phase may form other types of SDs 
\citep[][ see also Sect. \ref{sec:blaps}]{2023ApJ...959...24Z}. 

\subsection{Magnetic field generation in WD--WD mergers}\label{sect:magnetic_merger}

The merging process is predicted to dramatically amplify magnetic fields via a small-scale dynamo in the rapidly rotating merger product \citep{2013ApJ...773..136J}.
Recently, \cite{2024arXiv240702566P} found that also a large-scale dynamo may occur in a merger of two He-WDs producing a large-scale ordered azimuthal field by  magneto-rotational instability. Hence, SDs formed through a merger should be magnetic unless the magnetic fields dissipate quickly.

The simulations of \cite{2024arXiv240702566P} indicate that this is, indeed, the case for unequal mass mergers, where He burning starts in a shell generating a convective zone that erases the strong-order magnetic field. On the contrary, equal-mass mergers start He burning in the core and the high magnetic field will persist until long after the merger, which might explain the small fraction of SDs  found to be strongly magnetic.  Accordingly, the mass ratio determines whether the SD resulting from a He-WD merger is magnetic or not.

\begin{SCfigure}
\centering
\includegraphics[width=0.65\textwidth]{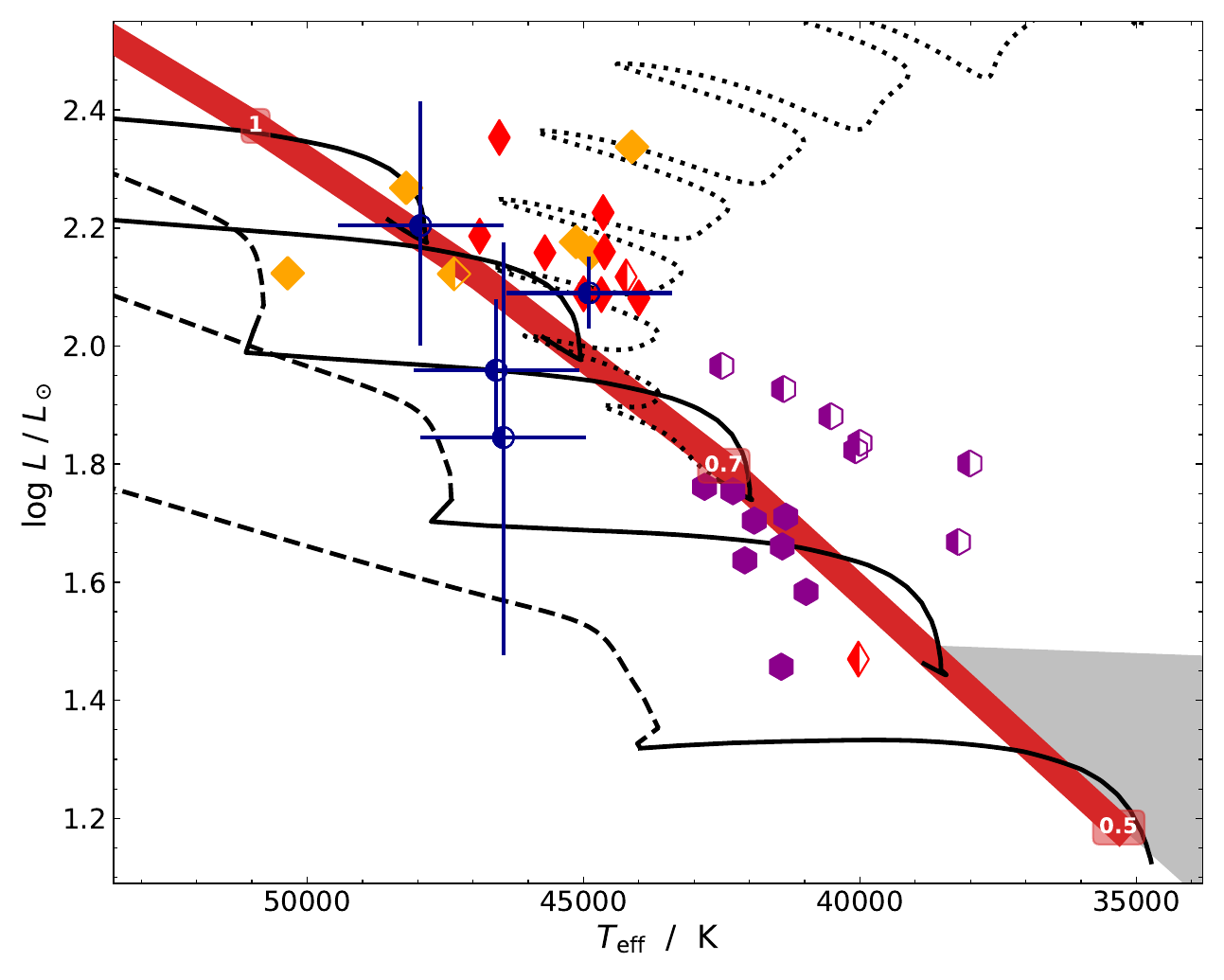} 
\hspace*{0.011\textwidth}
\caption{Pre- to post-merger evolution \cite[black lines, ][]{2021MNRAS.504.2670Y} in the HRD 
for remnant masses of 0.85, 0.75, 0.65, 0.55, and 0.45 M$_\odot$, with the core and shell helium burning phases shown as solid and dashed lines, respectively. The dotted line represents the pre-helium MS evolution from a 0.65 M$_\odot$ track; the red band shows the HeMS with masses (M$_\odot$) labelled and the gray band the EHB. The observed distribution of He-rich hot subdwarf stars \citep[see][]{2016PASP..128h2001H} is also shown. Extremely He-rich stars are marked by filled symbols and intermediately He-rich stars by half filled, half open symbols, N-rich ones by purple hexagons, C-rich by orange diamonds, C\&N-rich by red diamonds. Magnetic SDs are shown as blue circles (with error bars). Courtesy M. Dorsch. 
}\label{fig:sdO_distribution}
\end{SCfigure}

\begin{SCfigure}
\centering
\includegraphics[width=0.65\textwidth]{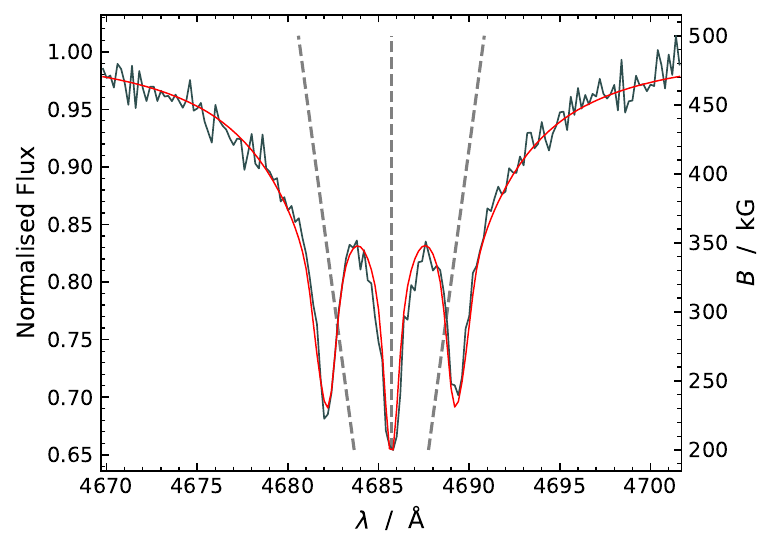} 
\hspace*{0.011\textwidth}       
\caption{Zeeman-split He {\sc{ii}} 4686$\AA$\ line in the spectrum of the magnetic HesdO Gaia DR2 5694207034772278400 (gray) with best fit model spectrum (red). The dependence of the Zeeman components on the magnetic field strength is illustrated by dashed lines \citep[modified from Fig. 1 of][courtesy M. Dorsch]{2022A&A...658L...9D}.}\label{fig:zeeman}
\end{SCfigure}

\begin{SCfigure}
    \centering
\includegraphics[width=0.65\textwidth]{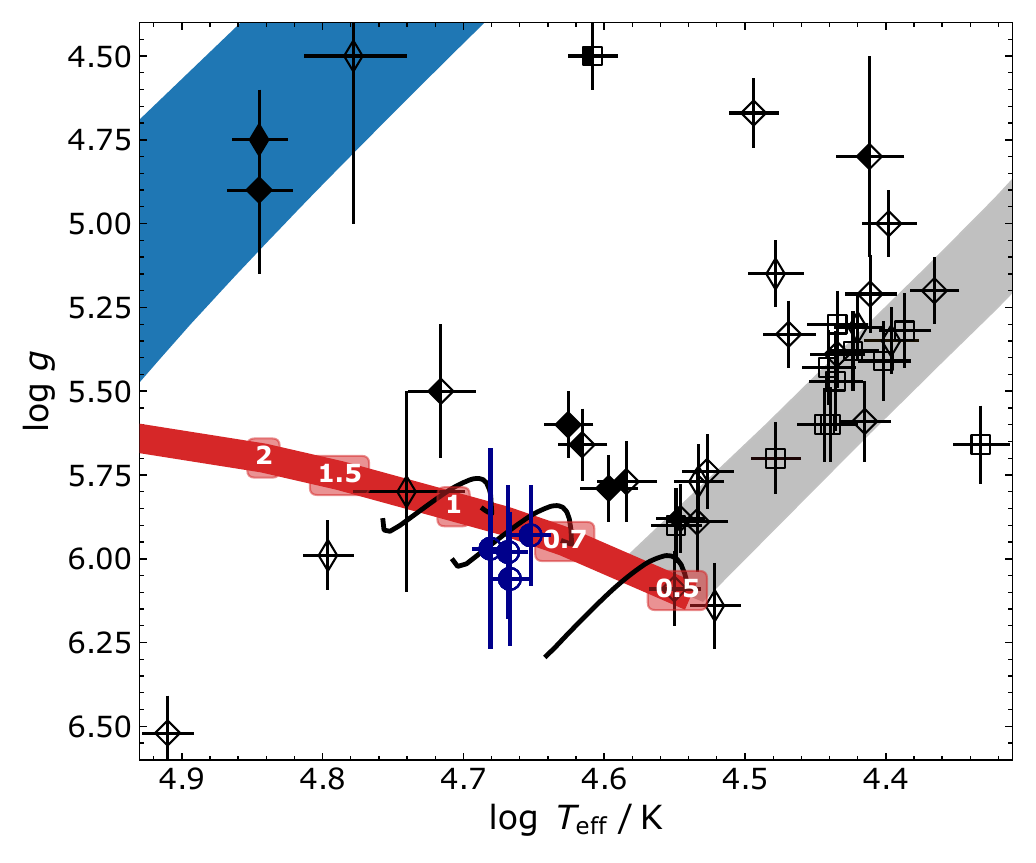}
\hspace*{0.011\textwidth}
\caption{Hot SDs probed for magnetic fields.  The four known magnetic HesdOs (blue circles) lie close to the HeMS at a high mass of $\approx$0.8 M$_\odot$. No evidence for the presence of a magnetic field has been found in any of the other stars. 
He-poor, eHe-, and iHe-rich stars are marked by open, filled, and half-filled, half open symbols, respectively. Black diamonds mark non-RV variable stars, and squares show RV variables  (close binaries with WD or MS companions).
The core helium burning phase of merger tracks of \cite{2021MNRAS.504.2670Y} 
are shown for
remnant masses of 0.45, 0.65, 0.85 M$_\odot$. The gray shaded region marks the location of the EHB  \citep[][solar metallicity]{1993ApJ...419..596D}, the blue shaded region marks the range of post-AGB tracks \citep[modified from Fig. 7 of] [courtesy M. Dorsch]{2022MNRAS.515.2496P}.}\label{fig:magnetic}
\end{SCfigure}

For a long time, a lot of effort has been invested in searching for strong magnetic fields in SDs in all relevant regions of the HRD 
as predicted to be created in mergers. However, no spectral evidence had been found until recently. Several spectropolarimetric searches targeting different spectral types failed to detect magnetic fields. Serendipitously, \cite{2022A&A...658L...9D} came across an iHesdO with a Zeeman-split spectrum
(see Fig. \ref{fig:zeeman}) indicating the presence of a magnetic field of 350 kG. Soon thereafter, three additional iHesdOs were found to display very similar field strengths as well as atmospheric parameters and helium abundances (see Fig. \ref{fig:sdO_distribution}). The pre- and post-merger evolution \citep[e.g.][]{2012MNRAS.419..452Z,2021MNRAS.504.2670Y} matches the observed position of the magnetic iHesdOs supporting the scenario that strongly magnetic subdwarfs result from 
mergers. Their position next to the HeMS suggests that they are more massive ($\approx$0.8 M$_\odot$) than normal SDs, but not necessarily more massive than non-magnetic He-sdOs {
 \citep[][and Fig. \ref{fig:magnetic}]{2022MNRAS.515.2496P}}.  

\subsection{New evidence for single hot subdwarfs to form from mergers}

Single hot SDs could result from WD mergers, preferentially forming He-rich objects, which is supported by recent discoveries.  
Two single He-rich sdOs have been reported, probably formed in an exceptional merger event. Those SDs exhibit a large fraction of helium-burning ash (C, O) on their surfaces \citep{2022MNRAS.511L..66W}. In the suggested merger of a He-WD and a C/O-WD, the C/O-WD must have been the less massive (larger) WD ($\approx$0.35 M$_\odot$) to transfer mass to the He-WD. Such low-mass C/O-WDs can form from RGB progenitor masses of $\approx$2.0 M$_\odot$, where He ignites in non-degenerate conditions \citep{2022MNRAS.511L..60M}.

\begin{SCfigure}
\centering
\includegraphics[width=0.5\textwidth]{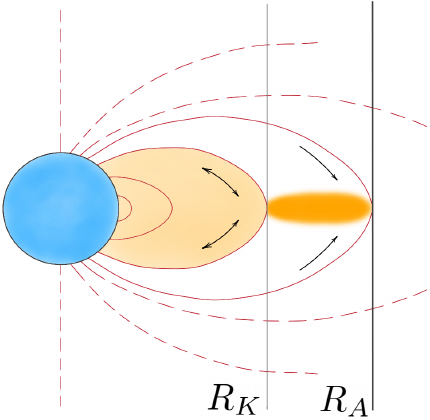}
\hspace*{0.011\textwidth}
\caption{Model for a rapidly spinning SD with a gas disk \citep{2021A&A...655A..43V}: The spinning SD (in blue) generates a bipolar magnetic field around it, which is not necessarily aligned with the spin. The wind material in the innermost region, within Kepler radius RK, has no centrifugal support and can fall freely back onto the star, thus forming a dynamical magnetosphere. The material between the Keplerian co-rotation radius RK and the Alfvén radius RA has centrifugal support. It cannot fall back onto the star, thus forming a much denser rigidly rotating magnetosphere \citep[reproduced with permission by A\&A from ][Fig. 6; copyright ESO]{2021A&A...655A..43V}.}\label{fig:braking}
\end{SCfigure}

The angular momentum of a post-merger should be high and, therefore, the object should be rapidly rotating. 
A rapidly spinning sdB with a circumstellar gas disk \citep[J22564-5910,][]{2021A&A...655A..43V} is a promising candidate to be a relic of a merger that happened shortly before. In their model, a magnetic field induced by the rapidly spinning SD  allows a gas disk to remain between the Keplerian co-rotating and the Alfvén radii after the merger completed (see Fig. \ref{fig:braking}).

\section{Pulsations, rotation, and spin-orbit synchronization}\label{sect:pulsation}

Radial and non-radial pulsations have been discovered in SD stars. They are excited by an opacity mechanism which converts heat into mechanical work like a piston in a steam engine. An opacity-ionization cycle is the driver of the pulsation. A zone of partially ionized metals, in particular Fe and Ni, in sub-photospheric layers blocks the radiation from lower more transparent layers, heating the zone, and forcing it to expand. Thereafter, the radiation escapes and the zone cools again, contracts, and recreates the partially ionized high opacity layer, which starts the cycle once more. Whether the oscillations are damped depends on the detailed structure of the envelope and defines instability strips in the HRD. To drive pulsations in SD stars, Fe and Ni have to be enriched via radiative levitation in the partially ionized zones.

Two types of SD pulsators can be distinguished, depending on the nature of the restoring force. The restoring force for p-modes is {
the pressure force and the buoyancy force} for g-modes
\citep[see][for a review]{2021RvMP...93a5001A}. The p-modes are seen in the hotter stars ($\approx$28,000 to 36,000 K), while g-modes are found at lower temperatures
(22,000 to 30,000 K). 
Observed periods are short (few min) for p-modes and longer for g-modes ($\approx$45--250 min). Many SD pulsators show both modes (hybrid pulsators), which are particularly useful for asteroseismology as g-modes penetrate deeper into the interior, while p-modes probe the outer layers, allowing the internal structure of a star to be investigated in great detail, in particular to study differential rotation, which has successfully been done for a few SDs in binaries with MS companions \citep{2024arXiv240717887S}. 

Both observation and analysis of light variations of pulsating SDs are demanding.  Precise time-series photometry of short cadence and long duration is required. Ground-based observational campaigns suffer from day-night gaps and, therefore, multi-site campaigns were organized. The \textit{Kepler} mission was a game changer because it provided LCs with unprecedented precision and short cadence (1 min), almost uninterrupted for months. This allowed the SD pulsation frequency spectrum to be resolved \citep[see][for reviews]{2021FrASS...8...19L,2021MNRAS.507.4178R}.
The analysis of p-mode pulsators benefits from the ultra-short cadence (20s) of \textit{TESS} LCs.  

Mode trapping is frequently observed.
The H/He and He/C-O transition regions in the interiors of SDs define cavities, where modes can be trapped. The trapped modes identified in the rich pulsation frequency spectra of pulsating SDs hold important information on the location of discontinuities
in the interior.

\subsection{Blue Large Amplitude Pulsators}\label{sec:blaps}

Recently, a new class of pulsator, so-called blue large amplitude pulsators \citep[BLAP, see][for a review]{2024arXiv240416089P}, is characterized by single-mode radial pulsations with typically saw-tooth shaped LCs similar to those of RR~Lyr stars. Spectroscopically, they share the ranges of temperature and He abundances with the SDs, but their gravities are typically lower,  forming two groups.
A low-gravity  one ($\log g$=4.2 to 4.7) with pulsation periods between 20--60 min and amplitudes of 0.2--0.4 mag, and a high-gravity one ($\log g$ = 5.3--5.7) with shorter periods (2--8 min) and lower amplitudes (0.05--0.2 mag).
Long-term observations indicate that the pulsation periods change by order of 10$^{-7}$ yr$^{-1}$. For most BLAPs the pulsation period is increasing, while for about a quarter it is found to decrease. 

Two rivaling scenarios for BLAP evolution have been proposed \citep[see] [for details]{2019ApJ...878L..35K}. The first scenario considers BLAPs to be He cores of giants stripped before the tip of the RGB was reached. Their helium core is inert, that is, they are the progenitors of low-mass ($\approx$0.3 M$_\odot$) WDs    
passing through the BLAP instability zone (see also Sect. \ref{sect:elm}). 
In the second scenario, BLAPs are considered as either progenitors or the progeny of helium burning EHB stars. Accordingly, BLAPs could be evolving towards or away from the EHB. Hence, BLAPs should have masses similar to EHB stars ($\approx$0.5 M$_\odot$).
Because the observed periods of low-order radial oscillations show better agreement with their models of pre-WD models than with pre-/post-EHB models, \cite{2019ApJ...878L..35K} prefer the former scenario. 

Recently, \cite{2023ApJ...959...24Z} investigated models of 
mergers of HeWD with MS stars, which evolve through the observed regime in the HRD occupied by BLAPS after He-shell burning started but before helium-core burning is fully established. The pulsations are excited via the iron bump opacity as in the SD pulsators and the models' pulsation properties match the observations.
Helium-shell flashes introduce periods of expansion and contraction, which are  consistent with observed rates of period change. Hence, the HeWD+MS merger model is a viable formation channel, relevant to explaining single BLAPs.

\begin{figure}
    \includegraphics[width=\textwidth]{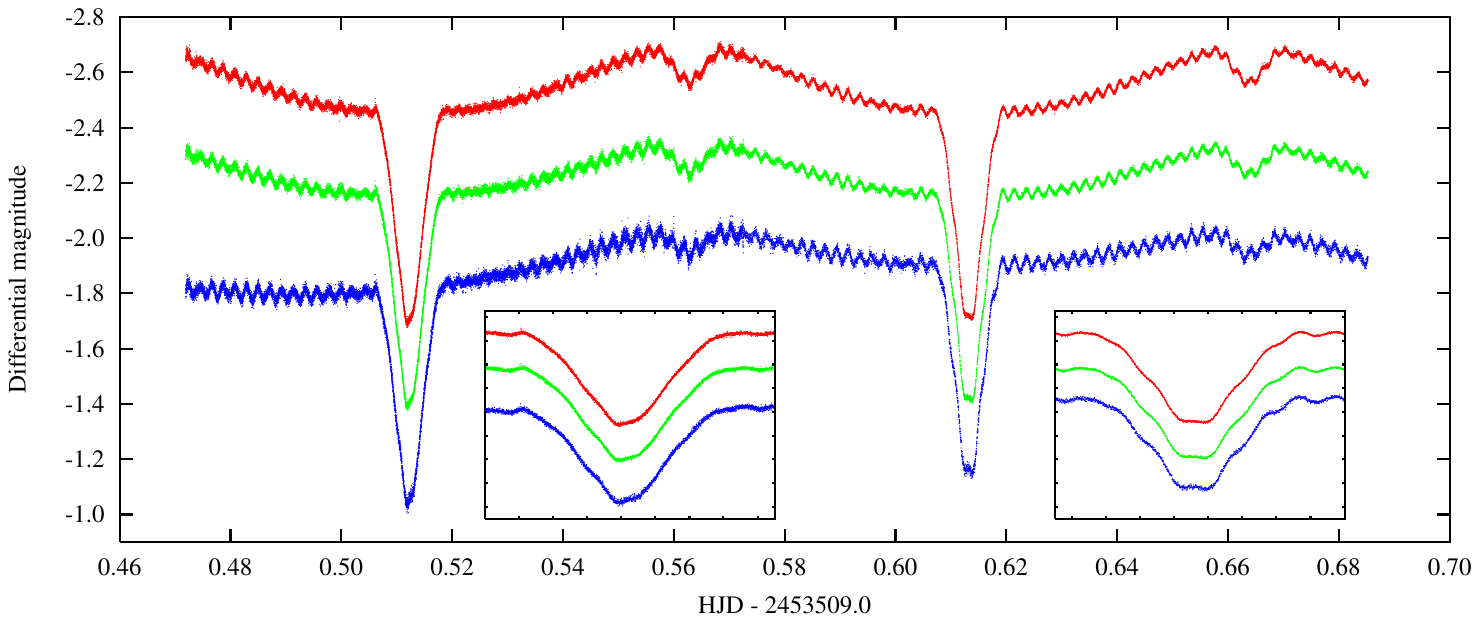}

    \vspace*{-4cm}
\caption{Light curves of the eclipsing sdB star NY Vir, simultaneously recorded in three filters. In addition to variations due to the reflected light from the secondary and the primary eclipses, multi-periodic pulsations are obvious 
\citep{2007A&A...471..605V}. {
Two consecutive primary eclipses, are shown in the insets (enlarged). Beating of the pulsation modes cause the differences between those primary eclipses.} NY~Vir allows mass determination in two independent ways: from RV and LC curve as well as from asteroseismology resulting in consistent values 
\citep[modified from Fig. 1 of][]{2007A&A...471..605V}. }
    \label{fig:ny_vir}
\end{figure}

\subsection{Probing stellar rotation via pulsations}

The analysis of spectral lines allows the measurement of the projected rotational velocity v$_\textrm{rot}$\,$\sin{i}$ of a star. In single SDs, v$_\textrm{rot}$\,$\sin{i}$ is small, at or below the detection limit of a few km\,s$^{-1}$ of high-resolution spectra. In close binaries, tidal forces can spin-up rotation, which may lead to alignment of rotation axes and synchronization of the rotation with the binary orbit. As the tidal forces increase in strength with decreasing orbital period, we may expect synchronization in the shortest period systems. Indeed, the SDs in those systems are spun-up to $\approx$200 km\,s$^{-1}$ in the most extreme case \citep{2024NatAs...8..491L}.
Model predictions for synchronization of stars with radiative envelopes, such as SD stars, are very uncertain. Therefore, empirical relations between the rotational and orbital periods have to be established, which can be done for slowly rotating stars using asteroseismology. 
Long-term LCs of pulsating stars allow us to study stellar rotation, which lifts the azimuthal degeneracy and produces equally spaced frequency multiplets. From the mean multiplet spacing, the long rotational periods of 20 -- 300 d have been derived  for single stars \citep[see][for a review]{2022MNRAS.511.2201S}, which is too slow to be detectable from spectral line broadening. 
As the tidal forces in SD binaries increase in strength with decreasing orbital period, we may expect synchronization preferentially in the shortest period systems. \citet{2022MNRAS.511.2201S} found that synchronization can be established in close SDs with orbital periods of about a quarter of a day or less, while in wider systems the SD rotation is not locked to the orbit. Models of convective coupling based on mixing length theory fail to reach synchronous rotation even for enlarged cores \citep{2019MNRAS.485.2889P},
in particular for the best studied case (NY Vir), calling for improvements to convection modelling. {
Recently, \citet{2024ApJ...975....1M} found that the tidal excitation of internal gravity waves in canonical mass sdB binaries can lead to tidal synchronization within the sdB lifetime for binary periods below 0.2d, consistent with observations.}

\begin{figure}
    \centering
    \begin{minipage}{0.59\textwidth}
    \includegraphics[width=\textwidth]{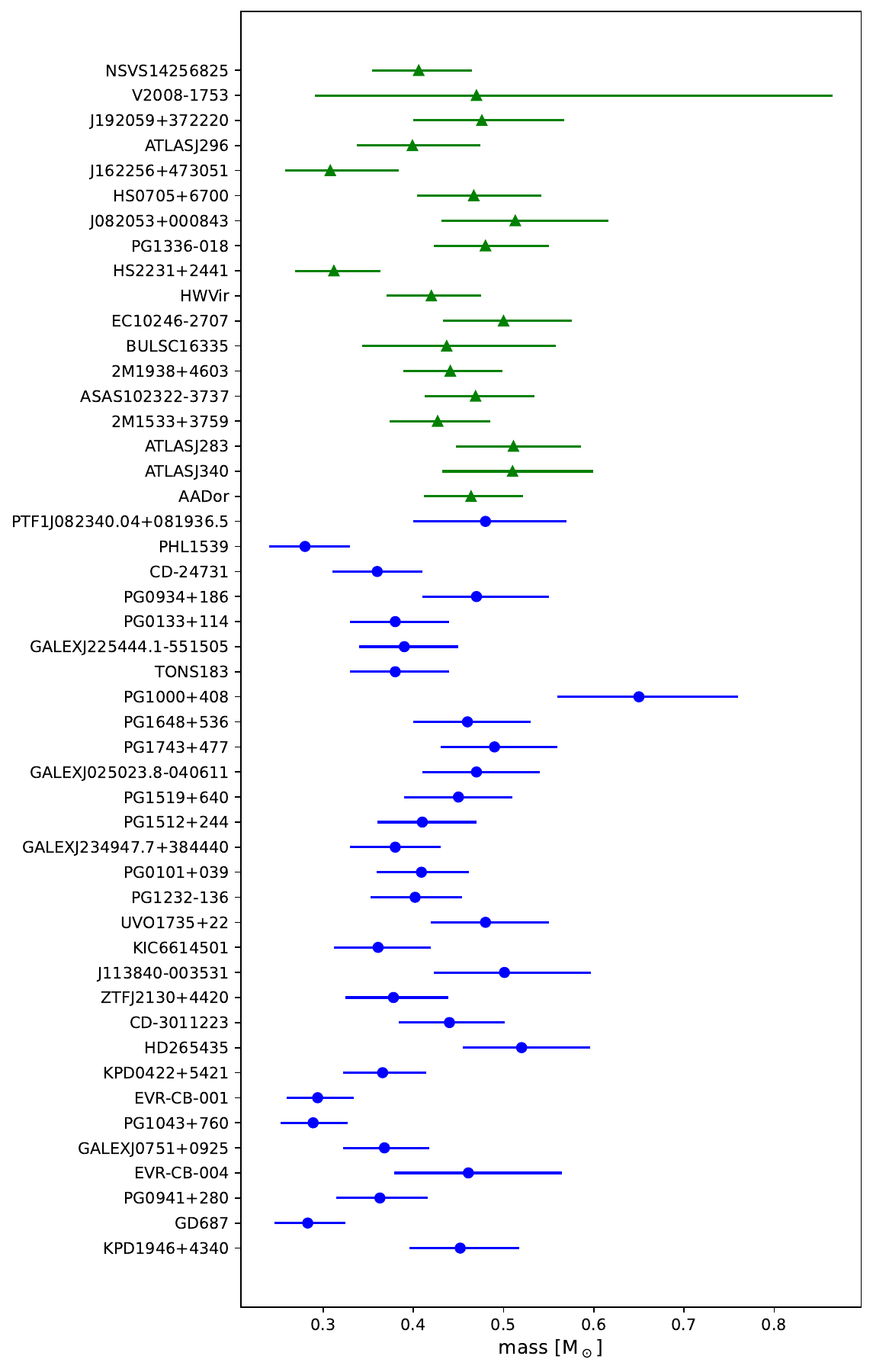}
    \end{minipage}
    \hfill
    \begin{minipage}{0.39\textwidth}
    \includegraphics[width=\textwidth]{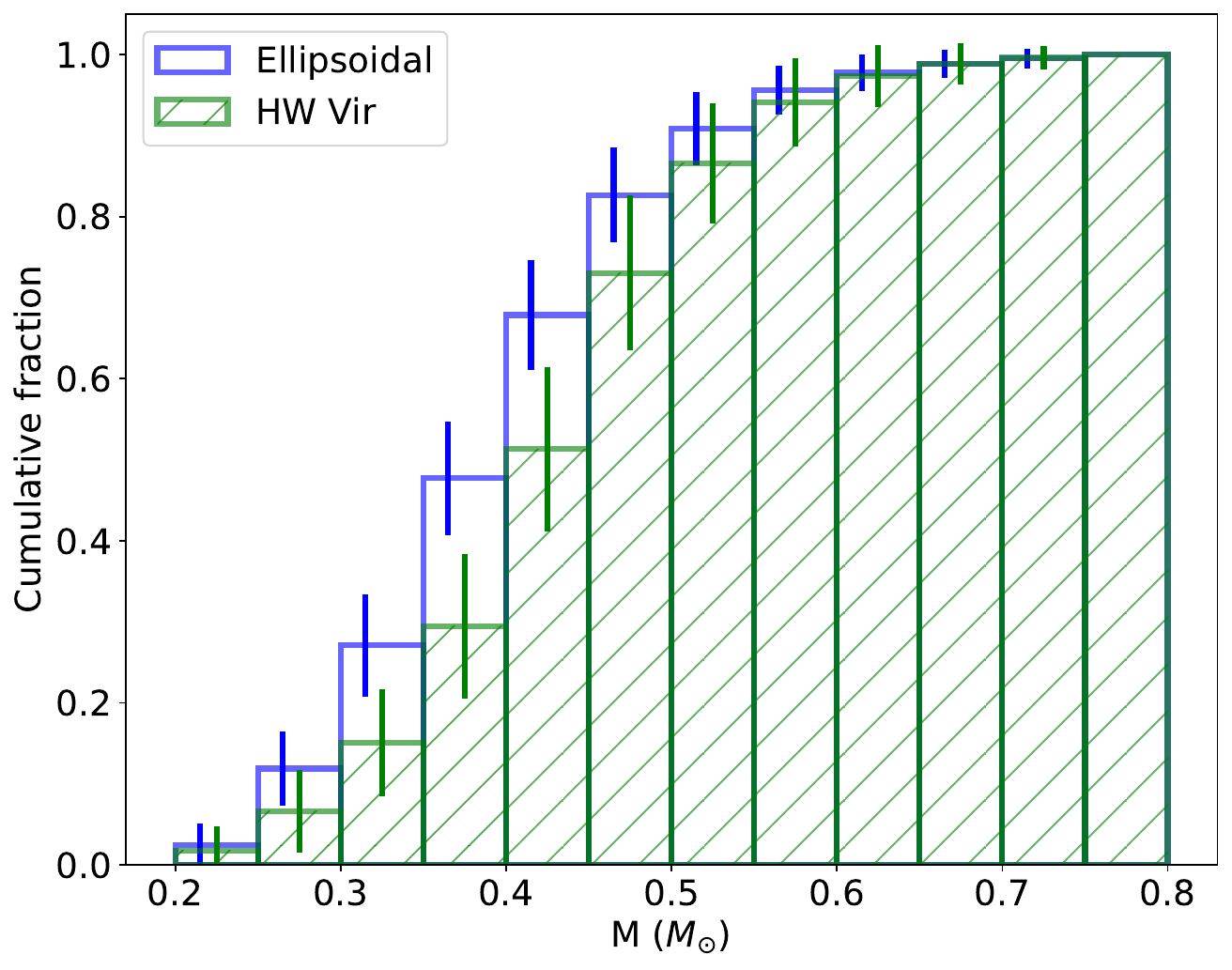}
        \caption{{\bf Left:} Subdwarf masses determined by combining the results of spectroscopic and SED analysis with Gaia parallaxes. Blue dots mark systems showing ellipsoidal deformation, green triangles mark HW Vir systems. {\bf Right:}  Cumulative distribution of the masses for the types of systems \citep[modified from Figs. 11 and 13 of][courtesy V. Schaffenroth]{2022A&A...666A.182S}. 
    }
    \vspace*{7.15cm}
    \end{minipage}
    \label{fig:masses}
\end{figure}

\section{Stellar parameters}\label{sect:parameters}

Typically, radii, luminosities, and masses are derived from observations of eclipsing, double-lined binaries. Among SD binaries, there are very  few suitable systems. Asteroseismology is another promising way for mass determination, because of the rich pulsation frequency spectra observed for many SD pulsators. Pulsating SD components in eclipsing binaries \citep{2022MNRAS.511.2201S} allow us to combine the two complementary methods. A well studied short-period sdB+MS binary is NY~Vir (see Fig. \ref{fig:ny_vir}), a p-mode pulsator in a tidally locked orbit with an MS star. The analysis of p-modes results in a sdB mass of 
M={
0.471}
$\pm$0.006 M$_\odot$ \citep{2013A&A...553A..97V}, very close to the canonical value 
\citep[0.47 M$_\odot$,][]{1993ApJ...419..596D}, 
whereas the analyses of the LC/RV curves by \cite{2007A&A...471..605V}  resulted in three equally probable solutions 
of 0.530, 0.466, and 0.389 M$_\odot$. Because of the scarcity of pulsating SDs in binaries, such mass determinations remained limited to a few SD stars.

\textit{Gaia} parallaxes provide a new opportunity to derive stellar parameters when \textit{Gaia} data are combined with multi-band photometry and quantitative spectral analyses.   
Stellar radii $R$ can readily be derived by combining angular diameters with \textit{Gaia} based distances. The angular diameters can be derived from spectral energy distributions (SEDs) by fitting multi-filter photometry with model spectra when interstellar reddening and extinction are taken into account. These model SED  fits allow us to estimate T$_\textrm{eff}$, from which the luminosity can be derived, and
to construct the HRD. The mass of a star is its most fundamental parameter, as determined using Newton's law of gravity, $M = g R^2 / G$ where $G$ is the gravitational constant. This determination requires an accurate measurement of the star's surface gravity $g$, which is obtained through quantitative spectral analysis. 

In a comprehensive study, \cite{2022A&A...666A.182S} derived masses of 7 ellipsoidal variables (SD+WD) using this method, compared them to the results derived from RV+LC analyses and found them to agree well. 
Masses of 18 eclipsing, reflection effect systems (sdB+MS, HW Vir-type) and 30 ellipsoidal variable SD+WD systems were calculated. The mass distribution of the former peaks at 0.46$^{+0.08}_{-0.12}$ M$_\odot$, that is, at the canonical mass \citep[0.47 M$_\odot$,][]{1993ApJ...419..596D}, while the masses of the latter are somewhat lower (0.38$^{+0.12}_{-0.08}$ M$_\odot$) on average (see Fig.\ref{fig:masses}, left hand panel). The cumulative distribution (see Fig. \ref{fig:masses}, right hand panel) provides
some evidence that the SD masses in close binaries with WDs, indeed, are lighter than those with MS stars.

\section{Conclusion and outlook}\label{sect:conclusion}

Hot subdwarf stars are important laboratories for understanding close-binary evolution through CE phases, as well as stable RLOF, and merger of binary components to form single SDs. Major uncertainties in binary evolution theory are the physical parameters of CE formation, evolution, and ejection as well as for the relevant merger process of WD+WD and WD+MS binaries. The mass-ratio threshold is important for understanding mass transfer stability. Whether mass transfer is conservative or not has also to be checked from observations.   
The observations of various new SD subclasses include magnetic He-rich sdOBs, SDs polluted by ashes of 3$\alpha$ processed material, an astrometric binary with a massive WD/NS companion in a long-period orbit, candidate SN~Ia progenitors, and blue large-amplitude pulsators. Asteroseismology as well as classical analysis of RV and LC curves allow us to derive the stellar parameters, important to constrain evolutionary models, but are limited to a few examples up to now. 
High-precision parallaxes from the \textit{Gaia} mission allow the determination of radii and luminosities when combined with multi-band photometry for numerous SD binaries of all kinds, as well as for single stars. The most fundamental parameter is the stellar mass, which can be derived from Newton's law but requires the surface gravity to be known to high precision to realize the potential of \textit{Gaia}'s parallax precision. Sophisticated quantitative spectral analyses of high-fidelity spectra are required to achieve sufficient accuracy of the gravity determinations. Large spectroscopic surveys, like ESO/4MOST, will provide the treasure trove of data for this purpose.     
The key to testing BPS models lies with volume limited samples, which avoid the biases inherent to flux limited samples. Volume-complete samples can be derived with 
\textit{Gaia} parallaxes. For SD stars, such a sample has already been established, which contains 305 SDs and is complete to 95\% at a distance of 500pc \citep{2024A&A...686A..25D}.
The analysis of follow-up observations is underway to provide us with a complete census of the characteristics of many 
SD subclasses in the solar neighborhood. 

\begin{ack}[Acknowledgments]

Many comments by Matti Dorsch, Stephan Geier, Thomas Kupfer, Tony Lynas-Gray, Veronika Schaffenroth, {
Valerie Van Grootel, and Lev Yungelson,} are gratefully acknowledged. 
I am much obliged to Harry Dawson, Matti Dorsch, and Lennard Kufleitner for providing me with their unpublished figures. I thank Matti Dorsch, Zhenxin Lei, Veronika Schaffenroth, and Xianfei Zhang for modifying published figures used in this manuscript and thank Aakash Bhat, Lennard Kufleitner, and Sebastian Weich for proofreading. 
\end{ack}

\bibliographystyle{Harvard}
\bibliography{els_reference_sd}

\begin{thebibliography*}{85}
\providecommand{\bibtype}[1]{}
\providecommand{\natexlab}[1]{#1}
{\catcode`\|=0\catcode`\#=12\catcode`\@=11\catcode`\\=12
|immediate|write|@auxout{\expandafter\ifx\csname
  natexlab\endcsname\relax\gdef\natexlab#1{#1}\fi}}
\renewcommand{\url}[1]{{\tt #1}}
\providecommand{\urlprefix}{URL }
\expandafter\ifx\csname urlstyle\endcsname\relax
  \providecommand{\doi}[1]{doi:\discretionary{}{}{}#1}\else
  \providecommand{\doi}{doi:\discretionary{}{}{}\begingroup
  \urlstyle{rm}\Url}\fi
\providecommand{\bibinfo}[2]{#2}
\providecommand{\eprint}[2][]{\url{#2}}

\bibtype{Article}%
\bibitem[{Aerts}(2021)]{2021RvMP...93a5001A}
\bibinfo{author}{{Aerts} C} (\bibinfo{year}{2021}), \bibinfo{month}{Jan.}
\bibinfo{title}{{Probing the interior physics of stars through
  asteroseismology}}.
\bibinfo{journal}{{\em Reviews of Modern Physics}} \bibinfo{volume}{93}
  (\bibinfo{number}{1}), \bibinfo{eid}{015001}.
  \bibinfo{doi}{\doi{10.1103/RevModPhys.93.015001}}.
\eprint{1912.12300}.

\bibtype{Article}%
\bibitem[{Barlow} et al.(2012)]{2012ApJ...753..101B}
\bibinfo{author}{{Barlow} BN}, \bibinfo{author}{{Wade} RA} and
  \bibinfo{author}{{Liss} SE} (\bibinfo{year}{2012}), \bibinfo{month}{Jul.}
\bibinfo{title}{{The R{\o}mer Delay and Mass Ratio of the sdB+dM Binary 2M
  1938+4603 from Kepler Eclipse Timings}}.
\bibinfo{journal}{{\em A\&A}} \bibinfo{volume}{753} (\bibinfo{number}{2}),
  \bibinfo{eid}{101}. \bibinfo{doi}{\doi{10.1088/0004-637X/753/2/101}}.
\eprint{1204.3321}.

\bibtype{Article}%
\bibitem[{Barlow} et al.(2024)]{2024A&A...686A.126B}
\bibinfo{author}{{Barlow} BN}, \bibinfo{author}{{Kupfer} T},
  \bibinfo{author}{{Smith} BA}, \bibinfo{author}{{Schaffenroth} V} and
  \bibinfo{author}{{Parker} I} (\bibinfo{year}{2024}), \bibinfo{month}{Jun.}
\bibinfo{title}{{Hot subdwarfs in close binaries observed from space. III.
  Reflection effect asymmetry induced by relativistic beaming}}.
\bibinfo{journal}{{\em A\&A}} \bibinfo{volume}{686}, \bibinfo{eid}{A126}.
  \bibinfo{doi}{\doi{10.1051/0004-6361/202348413}}.
\eprint{2402.13396}.

\bibtype{Article}%
\bibitem[{Battich} et al.(2023)]{2023A&A...680L..13B}
\bibinfo{author}{{Battich} T}, \bibinfo{author}{{Miller Bertolami} MM},
  \bibinfo{author}{{Serenelli} AM}, \bibinfo{author}{{Justham} S} and
  \bibinfo{author}{{Weiss} A} (\bibinfo{year}{2023}), \bibinfo{month}{Dec.}
\bibinfo{title}{{A self-synthesized origin for heavy metals in hot subdwarf
  stars}}.
\bibinfo{journal}{{\em A\&A}} \bibinfo{volume}{680}, \bibinfo{eid}{L13}.
  \bibinfo{doi}{\doi{10.1051/0004-6361/202348157}}.
\eprint{2311.04700}.

\bibtype{Article}%
\bibitem[{Chen} et al.(2024)]{2024PrPNP.13404083C}
\bibinfo{author}{{Chen} X}, \bibinfo{author}{{Liu} Z} and
  \bibinfo{author}{{Han} Z} (\bibinfo{year}{2024}), \bibinfo{month}{Jan.}
\bibinfo{title}{{Binary stars in the new millennium}}.
\bibinfo{journal}{{\em Progress in Particle and Nuclear Physics}}
  \bibinfo{volume}{134}, \bibinfo{eid}{104083}.
  \bibinfo{doi}{\doi{10.1016/j.ppnp.2023.104083}}.
\eprint{2311.11454}.

\bibtype{Article}%
\bibitem[{Culpan} et al.(2022)]{2022A&A...662A..40C}
\bibinfo{author}{{Culpan} R}, \bibinfo{author}{{Geier} S},
  \bibinfo{author}{{Reindl} N}, \bibinfo{author}{{Pelisoli} I},
  \bibinfo{author}{{Gentile Fusillo} N} and  \bibinfo{author}{{Vorontseva} A}
  (\bibinfo{year}{2022}), \bibinfo{month}{Jun.}
\bibinfo{title}{{The population of hot subdwarf stars studied with Gaia. IV.
  Catalogues of hot subluminous stars based on Gaia EDR3}}.
\bibinfo{journal}{{\em A\&A}} \bibinfo{volume}{662}, \bibinfo{eid}{A40}.
  \bibinfo{doi}{\doi{10.1051/0004-6361/202243337}}.
\eprint{2203.07938}.

\bibtype{Article}%
\bibitem[{Dawson} et al.(2024)]{2024A&A...686A..25D}
\bibinfo{author}{{Dawson} H}, \bibinfo{author}{{Geier} S},
  \bibinfo{author}{{Heber} U}, \bibinfo{author}{{Pelisoli} I},
  \bibinfo{author}{{Dorsch} M}, \bibinfo{author}{{Schaffenroth} V},
  \bibinfo{author}{{Reindl} N}, \bibinfo{author}{{Culpan} R},
  \bibinfo{author}{{Pritzkuleit} M}, \bibinfo{author}{{Vos} J},
  \bibinfo{author}{{Soemitro} AA}, \bibinfo{author}{{Roth} MM},
  \bibinfo{author}{{Schneider} D}, \bibinfo{author}{{Uzundag} M},
  \bibinfo{author}{{Vu{\v{c}}kovi{\'c}} M}, \bibinfo{author}{{Antunes Amaral}
  L}, \bibinfo{author}{{Istrate} AG}, \bibinfo{author}{{Justham} S},
  \bibinfo{author}{{{\O}stensen} RH}, \bibinfo{author}{{Telting} JH},
  \bibinfo{author}{{Djupvik} AA}, \bibinfo{author}{{Raddi} R},
  \bibinfo{author}{{Green} EM}, \bibinfo{author}{{Jeffery} CS},
  \bibinfo{author}{{Kepler} SO}, \bibinfo{author}{{Munday} J},
  \bibinfo{author}{{Steinmetz} T} and  \bibinfo{author}{{Kupfer} T}
  (\bibinfo{year}{2024}), \bibinfo{month}{Jun.}
\bibinfo{title}{{A 500 pc volume-limited sample of hot subluminous stars. I.
  Space density, scale height, and population properties}}.
\bibinfo{journal}{{\em A\&A}} \bibinfo{volume}{686}, \bibinfo{eid}{A25}.
  \bibinfo{doi}{\doi{10.1051/0004-6361/202348319}}.
\eprint{2403.15513}.

\bibtype{Article}%
\bibitem[{Deshmukh} et al.(2023)]{2023MNRAS.519..148D}
\bibinfo{author}{{Deshmukh} K}, \bibinfo{author}{{Kupfer} T},
  \bibinfo{author}{{Hakala} P}, \bibinfo{author}{{Bauer} EB},
  \bibinfo{author}{{Berdyugin} A}, \bibinfo{author}{{Bildsten} L},
  \bibinfo{author}{{Marsh} TR}, \bibinfo{author}{{Mereghetti} S} and
  \bibinfo{author}{{Piirola} V} (\bibinfo{year}{2023}), \bibinfo{month}{Feb.}
\bibinfo{title}{{Limiting the accretion disc light in two mass transferring hot
  subdwarf binaries}}.
\bibinfo{journal}{{\em MNRAS}} \bibinfo{volume}{519} (\bibinfo{number}{1}):
  \bibinfo{pages}{148--156}. \bibinfo{doi}{\doi{10.1093/mnras/stac3492}}.
\eprint{2211.12502}.

\bibtype{Article}%
\bibitem[{Deshmukh} et al.(2024)]{2024MNRAS.527.2072D}
\bibinfo{author}{{Deshmukh} K}, \bibinfo{author}{{Bauer} EB},
  \bibinfo{author}{{Kupfer} T} and  \bibinfo{author}{{Dorsch} M}
  (\bibinfo{year}{2024}), \bibinfo{month}{Jan.}
\bibinfo{title}{{Modelling the AM CVn and double detonation supernova
  progenitor binary system CD-30{\textdegree}11223}}.
\bibinfo{journal}{{\em MNRAS}} \bibinfo{volume}{527} (\bibinfo{number}{2}):
  \bibinfo{pages}{2072--2082}. \bibinfo{doi}{\doi{10.1093/mnras/stad3288}}.
\eprint{2310.01293}.

\bibtype{Article}%
\bibitem[{Dorman} et al.(1993)]{1993ApJ...419..596D}
\bibinfo{author}{{Dorman} B}, \bibinfo{author}{{Rood} RT} and
  \bibinfo{author}{{O'Connell} RW} (\bibinfo{year}{1993}),
  \bibinfo{month}{Dec.}
\bibinfo{title}{{Ultraviolet Radiation from Evolved Stellar Populations. I.
  Models}}.
\bibinfo{journal}{{\em ApJ}} \bibinfo{volume}{419}: \bibinfo{pages}{596}.
  \bibinfo{doi}{\doi{10.1086/173511}}.
\eprint{astro-ph/9311022}.

\bibtype{Article}%
\bibitem[{Dorsch} et al.(2020)]{2020A&A...643A..22D}
\bibinfo{author}{{Dorsch} M}, \bibinfo{author}{{Latour} M},
  \bibinfo{author}{{Heber} U}, \bibinfo{author}{{Irrgang} A},
  \bibinfo{author}{{Charpinet} S} and  \bibinfo{author}{{Jeffery} CS}
  (\bibinfo{year}{2020}), \bibinfo{month}{Nov.}
\bibinfo{title}{{Heavy-metal enrichment of intermediate He-sdOB stars: the
  pulsators Feige 46 and LS IV-14{\textdegree}116 revisited}}.
\bibinfo{journal}{{\em A\&A}} \bibinfo{volume}{643}, \bibinfo{eid}{A22}.
  \bibinfo{doi}{\doi{10.1051/0004-6361/202038859}}.
\eprint{2009.09032}.

\bibtype{Article}%
\bibitem[{Dorsch} et al.(2021)]{2021A&A...653A.120D}
\bibinfo{author}{{Dorsch} M}, \bibinfo{author}{{Jeffery} CS},
  \bibinfo{author}{{Irrgang} A}, \bibinfo{author}{{Woolf} V} and
  \bibinfo{author}{{Heber} U} (\bibinfo{year}{2021}), \bibinfo{month}{Sep.}
\bibinfo{title}{{EC 22536{\ensuremath{-}}5304: a lead-rich and metal-poor
  long-period binary}}.
\bibinfo{journal}{{\em A\&A}} \bibinfo{volume}{653}, \bibinfo{eid}{A120}.
  \bibinfo{doi}{\doi{10.1051/0004-6361/202141381}}.
\eprint{2107.06340}.

\bibtype{Article}%
\bibitem[{Dorsch} et al.(2022)]{2022A&A...658L...9D}
\bibinfo{author}{{Dorsch} M}, \bibinfo{author}{{Reindl} N},
  \bibinfo{author}{{Pelisoli} I}, \bibinfo{author}{{Heber} U},
  \bibinfo{author}{{Geier} S}, \bibinfo{author}{{Istrate} AG} and
  \bibinfo{author}{{Justham} S} (\bibinfo{year}{2022}), \bibinfo{month}{Feb.}
\bibinfo{title}{{Discovery of a highly magnetic He-sdO star from a
  double-degenerate binary merger}}.
\bibinfo{journal}{{\em A\&A}} \bibinfo{volume}{658}, \bibinfo{eid}{L9}.
  \bibinfo{doi}{\doi{10.1051/0004-6361/202142880}}.
\eprint{2201.08146}.

\bibtype{Article}%
\bibitem[{El-Badry}(2024)]{2024NewAR..9801694E}
\bibinfo{author}{{El-Badry} K} (\bibinfo{year}{2024}), \bibinfo{month}{Jun.}
\bibinfo{title}{{Gaia's binary star renaissance}}.
\bibinfo{journal}{{\em NewAR}} \bibinfo{volume}{98}, \bibinfo{eid}{101694}.
  \bibinfo{doi}{\doi{10.1016/j.newar.2024.101694}}.
\eprint{2403.12146}.

\bibtype{Article}%
\bibitem[{Fink} et al.(2007)]{2007A&A...476.1133F}
\bibinfo{author}{{Fink} M}, \bibinfo{author}{{Hillebrandt} W} and
  \bibinfo{author}{{R{\"o}pke} FK} (\bibinfo{year}{2007}),
  \bibinfo{month}{Dec.}
\bibinfo{title}{{Double-detonation supernovae of sub-Chandrasekhar mass white
  dwarfs}}.
\bibinfo{journal}{{\em A\&A}} \bibinfo{volume}{476} (\bibinfo{number}{3}):
  \bibinfo{pages}{1133--1143}. \bibinfo{doi}{\doi{10.1051/0004-6361:20078438}}.
\eprint{0710.5486}.

\bibtype{Article}%
\bibitem[{Ge} et al.(2022)]{2022ApJ...933..137G}
\bibinfo{author}{{Ge} H}, \bibinfo{author}{{Tout} CA}, \bibinfo{author}{{Chen}
  X}, \bibinfo{author}{{Kruckow} MU}, \bibinfo{author}{{Chen} H},
  \bibinfo{author}{{Jiang} D}, \bibinfo{author}{{Li} Z}, \bibinfo{author}{{Liu}
  Z} and  \bibinfo{author}{{Han} Z} (\bibinfo{year}{2022}),
  \bibinfo{month}{Jul.}
\bibinfo{title}{{The Common Envelope Evolution Outcome-A Case Study on Hot
  Subdwarf B Stars}}.
\bibinfo{journal}{{\em ApJ}} \bibinfo{volume}{933} (\bibinfo{number}{2}),
  \bibinfo{eid}{137}. \bibinfo{doi}{\doi{10.3847/1538-4357/ac75d3}}.
\eprint{2205.14256}.

\bibtype{Article}%
\bibitem[{Ge} et al.(2024)]{2024ApJ...961..202G}
\bibinfo{author}{{Ge} H}, \bibinfo{author}{{Tout} CA},
  \bibinfo{author}{{Webbink} RF}, \bibinfo{author}{{Chen} X},
  \bibinfo{author}{{Sarkar} A}, \bibinfo{author}{{Li} J}, \bibinfo{author}{{Li}
  Z}, \bibinfo{author}{{Zhang} L} and  \bibinfo{author}{{Han} Z}
  (\bibinfo{year}{2024}), \bibinfo{month}{Feb.}
\bibinfo{title}{{The Common Envelope Evolution Outcome. II.
  Short-orbital-period Hot Subdwarf B Binaries Reveal a Clear Picture}}.
\bibinfo{journal}{{\em ApJ}} \bibinfo{volume}{961} (\bibinfo{number}{2}),
  \bibinfo{eid}{202}. \bibinfo{doi}{\doi{10.3847/1538-4357/ad158e}}.
\eprint{2311.17304}.

\bibtype{Article}%
\bibitem[{Geier} et al.(2022)]{2022A&A...661A.113G}
\bibinfo{author}{{Geier} S}, \bibinfo{author}{{Dorsch} M},
  \bibinfo{author}{{Pelisoli} I}, \bibinfo{author}{{Reindl} N},
  \bibinfo{author}{{Heber} U} and  \bibinfo{author}{{Irrgang} A}
  (\bibinfo{year}{2022}), \bibinfo{month}{May}.
\bibinfo{title}{{Radial velocity variability and the evolution of hot subdwarf
  stars}}.
\bibinfo{journal}{{\em A\&A}} \bibinfo{volume}{661}, \bibinfo{eid}{A113}.
  \bibinfo{doi}{\doi{10.1051/0004-6361/202143022}}.
\eprint{2202.09608}.

\bibtype{Article}%
\bibitem[{Geier} et al.(2023)]{2023A&A...677A..11G}
\bibinfo{author}{{Geier} S}, \bibinfo{author}{{Dorsch} M},
  \bibinfo{author}{{Dawson} H}, \bibinfo{author}{{Pelisoli} I},
  \bibinfo{author}{{Munday} J}, \bibinfo{author}{{Marsh} TR},
  \bibinfo{author}{{Schaffenroth} V} and  \bibinfo{author}{{Heber} U}
  (\bibinfo{year}{2023}), \bibinfo{month}{Sep.}
\bibinfo{title}{{The first massive compact companion in a wide orbit around a
  hot subdwarf star}}.
\bibinfo{journal}{{\em A\&A}} \bibinfo{volume}{677}, \bibinfo{eid}{A11}.
  \bibinfo{doi}{\doi{10.1051/0004-6361/202346407}}.
\eprint{2305.03475}.

\bibtype{Article}%
\bibitem[{Geier} et al.(2024)]{2024arXiv240704479G}
\bibinfo{author}{{Geier} S}, \bibinfo{author}{{Heber} U},
  \bibinfo{author}{{Irrgang} A}, \bibinfo{author}{{Dorsch} M},
  \bibinfo{author}{{Bastian} A}, \bibinfo{author}{{Neunteufel} P},
  \bibinfo{author}{{Kupfer} T}, \bibinfo{author}{{Bloemen} S},
  \bibinfo{author}{{Kreuzer} S}, \bibinfo{author}{{M{\"o}ller} L},
  \bibinfo{author}{{Schindewolf} M}, \bibinfo{author}{{Schneider} D},
  \bibinfo{author}{{Ziegerer} E}, \bibinfo{author}{{Pelisoli} I},
  \bibinfo{author}{{Schaffenroth} V}, \bibinfo{author}{{Barlow} BN},
  \bibinfo{author}{{Raddi} R}, \bibinfo{author}{{Geier} SJ},
  \bibinfo{author}{{Reindl} N}, \bibinfo{author}{{Rauch} T},
  \bibinfo{author}{{Nemeth} P} and  \bibinfo{author}{{G{\"a}nsicke} BT}
  (\bibinfo{year}{2024}), \bibinfo{month}{Jul.}
\bibinfo{title}{{A spectroscopic and kinematic survey of fast hot subdwarfs}}.
\bibinfo{journal}{{\em arXiv e-prints}} ,
  \bibinfo{eid}{arXiv:2407.04479}\bibinfo{doi}{\doi{10.48550/arXiv.2407.04479}}.
\eprint{2407.04479}.

\bibtype{Article}%
\bibitem[{Gianninas} et al.(2016)]{2016ApJ...822L..27G}
\bibinfo{author}{{Gianninas} A}, \bibinfo{author}{{Curd} B},
  \bibinfo{author}{{Fontaine} G}, \bibinfo{author}{{Brown} WR} and
  \bibinfo{author}{{Kilic} M} (\bibinfo{year}{2016}), \bibinfo{month}{May}.
\bibinfo{title}{{Discovery of Three Pulsating, Mixed-atmosphere, Extremely
  Low-mass White Dwarf Precursors}}.
\bibinfo{journal}{{\em ApJL}} \bibinfo{volume}{822} (\bibinfo{number}{2}),
  \bibinfo{eid}{L27}. \bibinfo{doi}{\doi{10.3847/2041-8205/822/2/L27}}.
\eprint{1604.04621}.

\bibtype{Article}%
\bibitem[{Han} et al.(2003)]{2003MNRAS.341..669H}
\bibinfo{author}{{Han} Z}, \bibinfo{author}{{Podsiadlowski} P},
  \bibinfo{author}{{Maxted} PFL} and  \bibinfo{author}{{Marsh} TR}
  (\bibinfo{year}{2003}), \bibinfo{month}{May}.
\bibinfo{title}{{The origin of subdwarf B stars - II}}.
\bibinfo{journal}{{\em MNRAS}} \bibinfo{volume}{341} (\bibinfo{number}{2}):
  \bibinfo{pages}{669--691}.
  \bibinfo{doi}{\doi{10.1046/j.1365-8711.2003.06451.x}}.
\eprint{astro-ph/0301380}.

\bibtype{Article}%
\bibitem[{Heber}(2009)]{2009ARA&A..47..211H}
\bibinfo{author}{{Heber} U} (\bibinfo{year}{2009}), \bibinfo{month}{Sep.}
\bibinfo{title}{{Hot Subdwarf Stars}}.
\bibinfo{journal}{{\em ARAA}} \bibinfo{volume}{47} (\bibinfo{number}{1}):
  \bibinfo{pages}{211--251}.
  \bibinfo{doi}{\doi{10.1146/annurev-astro-082708-101836}}.

\bibtype{Article}%
\bibitem[{Heber}(2016)]{2016PASP..128h2001H}
\bibinfo{author}{{Heber} U} (\bibinfo{year}{2016}), \bibinfo{month}{Aug.}
\bibinfo{title}{{Hot Subluminous Stars}}.
\bibinfo{journal}{{\em PASP}} \bibinfo{volume}{128} (\bibinfo{number}{966}):
  \bibinfo{pages}{082001}.
  \bibinfo{doi}{\doi{10.1088/1538-3873/128/966/082001}}.
\eprint{1604.07749}.

\bibtype{Article}%
\bibitem[{Heber} et al.(2002)]{2002A&A...383..938H}
\bibinfo{author}{{Heber} U}, \bibinfo{author}{{Moehler} S},
  \bibinfo{author}{{Napiwotzki} R}, \bibinfo{author}{{Thejll} P} and
  \bibinfo{author}{{Green} EM} (\bibinfo{year}{2002}), \bibinfo{month}{Mar.}
\bibinfo{title}{{Resolving subdwarf B stars in binaries by HST imaging}}.
\bibinfo{journal}{{\em A\&A}} \bibinfo{volume}{383}: \bibinfo{pages}{938--951}.
  \bibinfo{doi}{\doi{10.1051/0004-6361:20020127}}.
\eprint{astro-ph/0201096}.

\bibtype{Article}%
\bibitem[{Hirsch} et al.(2005)]{2005A&A...444L..61H}
\bibinfo{author}{{Hirsch} HA}, \bibinfo{author}{{Heber} U},
  \bibinfo{author}{{O'Toole} SJ} and  \bibinfo{author}{{Bresolin} F}
  (\bibinfo{year}{2005}), \bibinfo{month}{Dec.}
\bibinfo{title}{{<ASTROBJ>US 708</ASTROBJ> - an unbound hyper-velocity
  subluminous O star}}.
\bibinfo{journal}{{\em A\&A}} \bibinfo{volume}{444} (\bibinfo{number}{3}):
  \bibinfo{pages}{L61--L64}. \bibinfo{doi}{\doi{10.1051/0004-6361:200500212}}.
\eprint{astro-ph/0511323}.

\bibtype{Article}%
\bibitem[{Hubeny} and {Lanz}(1995)]{1995ApJ...439..875H}
\bibinfo{author}{{Hubeny} I} and  \bibinfo{author}{{Lanz} T}
  (\bibinfo{year}{1995}), \bibinfo{month}{Feb.}
\bibinfo{title}{{Non-LTE Line-blanketed Model Atmospheres of Hot Stars. I.
  Hybrid Complete Linearization/Accelerated Lambda Iteration Method}}.
\bibinfo{journal}{{\em ApJ}} \bibinfo{volume}{439}: \bibinfo{pages}{875}.
  \bibinfo{doi}{\doi{10.1086/175226}}.

\bibtype{Article}%
\bibitem[{Irrgang} et al.(2021)]{2021A&A...650A.102I}
\bibinfo{author}{{Irrgang} A}, \bibinfo{author}{{Geier} S},
  \bibinfo{author}{{Heber} U}, \bibinfo{author}{{Kupfer} T},
  \bibinfo{author}{{El-Badry} K} and  \bibinfo{author}{{Bloemen} S}
  (\bibinfo{year}{2021}), \bibinfo{month}{Jun.}
\bibinfo{title}{{A proto-helium white dwarf stripped by a substellar companion
  via common-envelope ejection. Uncovering the true nature of a candidate
  hypervelocity B-type star}}.
\bibinfo{journal}{{\em A\&A}} \bibinfo{volume}{650}, \bibinfo{eid}{A102}.
  \bibinfo{doi}{\doi{10.1051/0004-6361/202038757}}.
\eprint{2007.03350}.

\bibtype{Article}%
\bibitem[{Jeffery}(2022)]{2022MNRAS.515..716J}
\bibinfo{author}{{Jeffery} CS} (\bibinfo{year}{2022}), \bibinfo{month}{Sep.}
\bibinfo{title}{{Spectrum synthesis for radially pulsating stars with shocked
  atmospheres}}.
\bibinfo{journal}{{\em MNRAS}} \bibinfo{volume}{515} (\bibinfo{number}{1}):
  \bibinfo{pages}{716--729}. \bibinfo{doi}{\doi{10.1093/mnras/stac1644}}.
\eprint{2206.05094}.

\bibtype{Article}%
\bibitem[{Ji} et al.(2013)]{2013ApJ...773..136J}
\bibinfo{author}{{Ji} S}, \bibinfo{author}{{Fisher} RT},
  \bibinfo{author}{{Garc{\'\i}a-Berro} E}, \bibinfo{author}{{Tzeferacos} P},
  \bibinfo{author}{{Jordan} G}, \bibinfo{author}{{Lee} D},
  \bibinfo{author}{{Lor{\'e}n-Aguilar} P}, \bibinfo{author}{{Cremer} P} and
  \bibinfo{author}{{Behrends} J} (\bibinfo{year}{2013}), \bibinfo{month}{Aug.}
\bibinfo{title}{{The Post-merger Magnetized Evolution of White Dwarf Binaries:
  The Double-degenerate Channel of Sub-Chandrasekhar Type Ia Supernovae and the
  Formation of Magnetized White Dwarfs}}.
\bibinfo{journal}{{\em ApJ}} \bibinfo{volume}{773} (\bibinfo{number}{2}),
  \bibinfo{eid}{136}. \bibinfo{doi}{\doi{10.1088/0004-637X/773/2/136}}.
\eprint{1302.5700}.

\bibtype{Article}%
\bibitem[{Kosakowski} et al.(2023{\natexlab{a}})]{2023ApJ...950..141K}
\bibinfo{author}{{Kosakowski} A}, \bibinfo{author}{{Brown} WR},
  \bibinfo{author}{{Kilic} M}, \bibinfo{author}{{Kupfer} T},
  \bibinfo{author}{{B{\'e}dard} A}, \bibinfo{author}{{Gianninas} A},
  \bibinfo{author}{{Ag{\"u}eros} MA} and  \bibinfo{author}{{Barrientos} M}
  (\bibinfo{year}{2023}{\natexlab{a}}), \bibinfo{month}{Jun.}
\bibinfo{title}{{The ELM Survey South. II. Two Dozen New Low-mass White Dwarf
  Binaries}}.
\bibinfo{journal}{{\em ApJ}} \bibinfo{volume}{950} (\bibinfo{number}{2}),
  \bibinfo{eid}{141}. \bibinfo{doi}{\doi{10.3847/1538-4357/acd187}}.
\eprint{2305.03079}.

\bibtype{Article}%
\bibitem[{Kosakowski} et al.(2023{\natexlab{b}})]{2023ApJ...959..114K}
\bibinfo{author}{{Kosakowski} A}, \bibinfo{author}{{Kupfer} T},
  \bibinfo{author}{{Bergeron} P} and  \bibinfo{author}{{Littenberg} TB}
  (\bibinfo{year}{2023}{\natexlab{b}}), \bibinfo{month}{Dec.}
\bibinfo{title}{{Electromagnetic Characterization of the LISA Verification
  Binary ZTF J0526+5934}}.
\bibinfo{journal}{{\em ApJ}} \bibinfo{volume}{959} (\bibinfo{number}{2}),
  \bibinfo{eid}{114}. \bibinfo{doi}{\doi{10.3847/1538-4357/ad0ce9}}.
\eprint{2307.00645}.

\bibtype{Article}%
\bibitem[{Kupfer} et al.(2019)]{2019ApJ...878L..35K}
\bibinfo{author}{{Kupfer} T}, \bibinfo{author}{{Bauer} EB},
  \bibinfo{author}{{Burdge} KB}, \bibinfo{author}{{Bellm} EC},
  \bibinfo{author}{{Bildsten} L}, \bibinfo{author}{{Fuller} J},
  \bibinfo{author}{{Hermes} J}, \bibinfo{author}{{Kulkarni} SR},
  \bibinfo{author}{{Prince} TA}, \bibinfo{author}{{van Roestel} J},
  \bibinfo{author}{{Dekany} R}, \bibinfo{author}{{Duev} DA},
  \bibinfo{author}{{Feeney} M}, \bibinfo{author}{{Giomi} M},
  \bibinfo{author}{{Graham} MJ}, \bibinfo{author}{{Kaye} S},
  \bibinfo{author}{{Laher} RR}, \bibinfo{author}{{Masci} FJ},
  \bibinfo{author}{{Porter} M}, \bibinfo{author}{{Riddle} R},
  \bibinfo{author}{{Shupe} DL}, \bibinfo{author}{{Smith} RM},
  \bibinfo{author}{{Soumagnac} MT}, \bibinfo{author}{{Szkody} P} and
  \bibinfo{author}{{Ward} C} (\bibinfo{year}{2019}), \bibinfo{month}{Jun.}
\bibinfo{title}{{A New Class of Large-amplitude Radial-mode Hot Subdwarf
  Pulsators}}.
\bibinfo{journal}{{\em ApJL}} \bibinfo{volume}{878} (\bibinfo{number}{2}),
  \bibinfo{eid}{L35}. \bibinfo{doi}{\doi{10.3847/2041-8213/ab263c}}.
\eprint{1906.00979}.

\bibtype{Article}%
\bibitem[{Kupfer} et al.(2020)]{2020ApJ...898L..25K}
\bibinfo{author}{{Kupfer} T}, \bibinfo{author}{{Bauer} EB},
  \bibinfo{author}{{Burdge} KB}, \bibinfo{author}{{Roestel} Jv},
  \bibinfo{author}{{Bellm} EC}, \bibinfo{author}{{Fuller} J},
  \bibinfo{author}{{Hermes} J}, \bibinfo{author}{{Marsh} TR},
  \bibinfo{author}{{Bildsten} L}, \bibinfo{author}{{Kulkarni} SR},
  \bibinfo{author}{{Phinney} ES}, \bibinfo{author}{{Prince} TA},
  \bibinfo{author}{{Szkody} P}, \bibinfo{author}{{Yao} Y},
  \bibinfo{author}{{Irrgang} A}, \bibinfo{author}{{Heber} U},
  \bibinfo{author}{{Schneider} D}, \bibinfo{author}{{Dhillon} VS},
  \bibinfo{author}{{Murawski} G}, \bibinfo{author}{{Drake} AJ},
  \bibinfo{author}{{Duev} DA}, \bibinfo{author}{{Feeney} M},
  \bibinfo{author}{{Graham} MJ}, \bibinfo{author}{{Laher} RR},
  \bibinfo{author}{{Littlefair} SP}, \bibinfo{author}{{Mahabal} AA},
  \bibinfo{author}{{Masci} FJ}, \bibinfo{author}{{Porter} M},
  \bibinfo{author}{{Reiley} D}, \bibinfo{author}{{Rodriguez} H},
  \bibinfo{author}{{Rusholme} B}, \bibinfo{author}{{Shupe} DL} and
  \bibinfo{author}{{Soumagnac} MT} (\bibinfo{year}{2020}),
  \bibinfo{month}{Jul.}
\bibinfo{title}{{A New Class of Roche Lobe-filling Hot Subdwarf Binaries}}.
\bibinfo{journal}{{\em ApJL}} \bibinfo{volume}{898} (\bibinfo{number}{1}),
  \bibinfo{eid}{L25}. \bibinfo{doi}{\doi{10.3847/2041-8213/aba3c2}}.
\eprint{2007.05349}.

\bibtype{Article}%
\bibitem[{Kupfer} et al.(2022)]{2022ApJ...925L..12K}
\bibinfo{author}{{Kupfer} T}, \bibinfo{author}{{Bauer} EB},
  \bibinfo{author}{{van Roestel} J}, \bibinfo{author}{{Bellm} EC},
  \bibinfo{author}{{Bildsten} L}, \bibinfo{author}{{Fuller} J},
  \bibinfo{author}{{Prince} TA}, \bibinfo{author}{{Heber} U},
  \bibinfo{author}{{Geier} S}, \bibinfo{author}{{Green} MJ},
  \bibinfo{author}{{Kulkarni} SR}, \bibinfo{author}{{Bloemen} S},
  \bibinfo{author}{{Laher} RR}, \bibinfo{author}{{Rusholme} B} and
  \bibinfo{author}{{Schneider} D} (\bibinfo{year}{2022}), \bibinfo{month}{Feb.}
\bibinfo{title}{{Discovery of a Double-detonation Thermonuclear Supernova
  Progenitor}}.
\bibinfo{journal}{{\em ApJL}} \bibinfo{volume}{925} (\bibinfo{number}{2}),
  \bibinfo{eid}{L12}. \bibinfo{doi}{\doi{10.3847/2041-8213/ac48f1}}.
\eprint{2110.11974}.

\bibtype{Article}%
\bibitem[{Kupfer} et al.(2024)]{2024ApJ...963..100K}
\bibinfo{author}{{Kupfer} T}, \bibinfo{author}{{Korol} V},
  \bibinfo{author}{{Littenberg} TB}, \bibinfo{author}{{Shah} S},
  \bibinfo{author}{{Savalle} E}, \bibinfo{author}{{Groot} PJ},
  \bibinfo{author}{{Marsh} TR}, \bibinfo{author}{{Le Jeune} M},
  \bibinfo{author}{{Nelemans} G}, \bibinfo{author}{{Pala} AF},
  \bibinfo{author}{{Petiteau} A}, \bibinfo{author}{{Ramsay} G},
  \bibinfo{author}{{Steeghs} D} and  \bibinfo{author}{{Babak} S}
  (\bibinfo{year}{2024}), \bibinfo{month}{Mar.}
\bibinfo{title}{{LISA Galactic Binaries with Astrometry from Gaia DR3}}.
\bibinfo{journal}{{\em ApJ}} \bibinfo{volume}{963} (\bibinfo{number}{2}),
  \bibinfo{eid}{100}. \bibinfo{doi}{\doi{10.3847/1538-4357/ad2068}}.
\eprint{2302.12719}.

\bibtype{Article}%
\bibitem[{Lagos} et al.(2020)]{2020MNRAS.499L.121L}
\bibinfo{author}{{Lagos} F}, \bibinfo{author}{{Schreiber} MR},
  \bibinfo{author}{{Parsons} SG}, \bibinfo{author}{{G{\"a}nsicke} BT} and
  \bibinfo{author}{{Godoy} N} (\bibinfo{year}{2020}), \bibinfo{month}{Dec.}
\bibinfo{title}{{Most EL CVn systems are inner binaries of hierarchical
  triples}}.
\bibinfo{journal}{{\em MNRAS}} \bibinfo{volume}{499} (\bibinfo{number}{1}):
  \bibinfo{pages}{L121--L125}. \bibinfo{doi}{\doi{10.1093/mnrasl/slaa164}}.
\eprint{2010.03507}.

\bibtype{Article}%
\bibitem[{Latour} et al.(2023)]{2023A&A...677A..86L}
\bibinfo{author}{{Latour} M}, \bibinfo{author}{{H{\"a}mmerich} S},
  \bibinfo{author}{{Dorsch} M}, \bibinfo{author}{{Heber} U},
  \bibinfo{author}{{Husser} TO}, \bibinfo{author}{{Kamman} S},
  \bibinfo{author}{{Dreizler} S} and  \bibinfo{author}{{Brinchmann} J}
  (\bibinfo{year}{2023}), \bibinfo{month}{Sep.}
\bibinfo{title}{{SHOTGLAS. II. MUSE spectroscopy of blue horizontal branch
  stars in the core of {\ensuremath{\omega}} Centauri and NGC6752}}.
\bibinfo{journal}{{\em A\&A}} \bibinfo{volume}{677}, \bibinfo{eid}{A86}.
  \bibinfo{doi}{\doi{10.1051/0004-6361/202346597}}.
\eprint{2306.14549}.

\bibtype{Article}%
\bibitem[{Ledda} et al.(2023)]{2023A&A...675A.184L}
\bibinfo{author}{{Ledda} S}, \bibinfo{author}{{Danielski} C} and
  \bibinfo{author}{{Turrini} D} (\bibinfo{year}{2023}), \bibinfo{month}{Jul.}
\bibinfo{title}{{The quest for Magrathea planets. I. Formation of
  second-generation exoplanets around double white dwarfs}}.
\bibinfo{journal}{{\em A\&A}} \bibinfo{volume}{675}, \bibinfo{eid}{A184}.
  \bibinfo{doi}{\doi{10.1051/0004-6361/202245827}}.
\eprint{2304.09204}.

\bibtype{Article}%
\bibitem[{Lei} et al.(2018)]{2018ApJ...868...70L}
\bibinfo{author}{{Lei} Z}, \bibinfo{author}{{Zhao} J},
  \bibinfo{author}{{N{\'e}meth} P} and  \bibinfo{author}{{Zhao} G}
  (\bibinfo{year}{2018}), \bibinfo{month}{Nov.}
\bibinfo{title}{{New Hot Subdwarf Stars Identified in Gaia DR2 with LAMOST DR5
  Spectra}}.
\bibinfo{journal}{{\em ApJ}} \bibinfo{volume}{868} (\bibinfo{number}{1}),
  \bibinfo{eid}{70}. \bibinfo{doi}{\doi{10.3847/1538-4357/aae82b}}.
\eprint{1810.09625}.

\bibtype{Article}%
\bibitem[{Lin} et al.(2024)]{2024NatAs...8..491L}
\bibinfo{author}{{Lin} J}, \bibinfo{author}{{Wu} C}, \bibinfo{author}{{Xiong}
  H}, \bibinfo{author}{{Wang} X}, \bibinfo{author}{{N{\'e}meth} P},
  \bibinfo{author}{{Han} Z}, \bibinfo{author}{{Li} J},
  \bibinfo{author}{{Elias-Rosa} N}, \bibinfo{author}{{Salmaso} I},
  \bibinfo{author}{{Filippenko} AV}, \bibinfo{author}{{Brink} TG},
  \bibinfo{author}{{Yang} Y}, \bibinfo{author}{{Chen} X},
  \bibinfo{author}{{Yan} S}, \bibinfo{author}{{Zhang} J},
  \bibinfo{author}{{Guo} S}, \bibinfo{author}{{Cai} Y}, \bibinfo{author}{{Mo}
  J}, \bibinfo{author}{{Xi} G}, \bibinfo{author}{{Liu} J},
  \bibinfo{author}{{Guo} J}, \bibinfo{author}{{Xia} Q},
  \bibinfo{author}{{Xiang} D}, \bibinfo{author}{{Li} G}, \bibinfo{author}{{Li}
  Z}, \bibinfo{author}{{Zheng} W}, \bibinfo{author}{{Zhang} J},
  \bibinfo{author}{{Liu} Q}, \bibinfo{author}{{Guo} F}, \bibinfo{author}{{Chen}
  L} and  \bibinfo{author}{{Li} W} (\bibinfo{year}{2024}),
  \bibinfo{month}{Apr.}
\bibinfo{title}{{A seven-Earth-radius helium-burning star inside a 20.5-min
  detached binary}}.
\bibinfo{journal}{{\em Nature Astronomy}} \bibinfo{volume}{8}:
  \bibinfo{pages}{491--503}. \bibinfo{doi}{\doi{10.1038/s41550-023-02188-2}}.
\eprint{2312.13612}.

\bibtype{Article}%
\bibitem[{Luo} et al.(2021)]{2021ApJS..256...28L}
\bibinfo{author}{{Luo} Y}, \bibinfo{author}{{N{\'e}meth} P},
  \bibinfo{author}{{Wang} K}, \bibinfo{author}{{Wang} X} and
  \bibinfo{author}{{Han} Z} (\bibinfo{year}{2021}), \bibinfo{month}{Oct.}
\bibinfo{title}{{Hot Subdwarf Atmospheric Parameters, Kinematics, and Origins
  Based on 1587 Hot Subdwarf Stars Observed in Gaia DR2 and LAMOST DR7}}.
\bibinfo{journal}{{\em ApJS}} \bibinfo{volume}{256} (\bibinfo{number}{2}),
  \bibinfo{eid}{28}. \bibinfo{doi}{\doi{10.3847/1538-4365/ac11f6}}.
\eprint{2107.09302}.

\bibtype{Article}%
\bibitem[{Luo} et al.(2024)]{2024ApJS..271...21L}
\bibinfo{author}{{Luo} Y}, \bibinfo{author}{{N{\'e}meth} P},
  \bibinfo{author}{{Wang} K} and  \bibinfo{author}{{Pan} Y}
  (\bibinfo{year}{2024}), \bibinfo{month}{Mar.}
\bibinfo{title}{{He-rich Hot Subdwarf Stars Observed in Gaia DR3 and LAMOST
  DR7: Carbon and Nitrogen Abundances and Kinematics}}.
\bibinfo{journal}{{\em ApJS}} \bibinfo{volume}{271} (\bibinfo{number}{1}),
  \bibinfo{eid}{21}. \bibinfo{doi}{\doi{10.3847/1538-4365/ad1ab2}}.

\bibtype{Article}%
\bibitem[{Lynas-Gray}(2021)]{2021FrASS...8...19L}
\bibinfo{author}{{Lynas-Gray} AE} (\bibinfo{year}{2021}), \bibinfo{month}{Apr.}
\bibinfo{title}{{Asteroseismic Observations of Hot Subdwarfs}}.
\bibinfo{journal}{{\em Frontiers in Astronomy and Space Sciences}}
  \bibinfo{volume}{8}, \bibinfo{eid}{19}.
  \bibinfo{doi}{\doi{10.3389/fspas.2021.576623}}.

\bibtype{Article}%
\bibitem[{Ma} and {Fuller}(2024)]{2024ApJ...975....1M}
\bibinfo{author}{{Ma} L} and  \bibinfo{author}{{Fuller} J}
  (\bibinfo{year}{2024}), \bibinfo{month}{Nov.}
\bibinfo{title}{{Tidal Spin-up of Subdwarf B Stars}}.
\bibinfo{journal}{{\em ApJ}} \bibinfo{volume}{975} (\bibinfo{number}{1}),
  \bibinfo{eid}{1}. \bibinfo{doi}{\doi{10.3847/1538-4357/ad7788}}.
\eprint{2408.16158}.

\bibtype{Article}%
\bibitem[{Masuda} et al.(2019)]{2019ApJ...881L...3M}
\bibinfo{author}{{Masuda} K}, \bibinfo{author}{{Kawahara} H},
  \bibinfo{author}{{Latham} DW}, \bibinfo{author}{{Bieryla} A},
  \bibinfo{author}{{Kunitomo} M}, \bibinfo{author}{{MacLeod} M} and
  \bibinfo{author}{{Aoki} W} (\bibinfo{year}{2019}), \bibinfo{month}{Aug.}
\bibinfo{title}{{Self-lensing Discovery of a 0.2 M $_{{\ensuremath{\odot}}}$
  White Dwarf in an Unusually Wide Orbit around a Sun-like Star}}.
\bibinfo{journal}{{\em ApJL}} \bibinfo{volume}{881} (\bibinfo{number}{1}),
  \bibinfo{eid}{L3}. \bibinfo{doi}{\doi{10.3847/2041-8213/ab321b}}.
\eprint{1907.07656}.

\bibtype{Article}%
\bibitem[{Maxted} et al.(2001)]{2001MNRAS.326.1391M}
\bibinfo{author}{{Maxted} PFL}, \bibinfo{author}{{Heber} U},
  \bibinfo{author}{{Marsh} TR} and  \bibinfo{author}{{North} RC}
  (\bibinfo{year}{2001}), \bibinfo{month}{Oct.}
\bibinfo{title}{{The binary fraction of extreme horizontal branch stars}}.
\bibinfo{journal}{{\em MNRAS}} \bibinfo{volume}{326} (\bibinfo{number}{4}):
  \bibinfo{pages}{1391--1402}.
  \bibinfo{doi}{\doi{10.1111/j.1365-2966.2001.04714.x}}.
\eprint{astro-ph/0103342}.

\bibtype{Article}%
\bibitem[{Maxted} et al.(2011)]{2011MNRAS.418.1156M}
\bibinfo{author}{{Maxted} PFL}, \bibinfo{author}{{Anderson} DR},
  \bibinfo{author}{{Burleigh} MR}, \bibinfo{author}{{Collier Cameron} A},
  \bibinfo{author}{{Heber} U}, \bibinfo{author}{{G{\"a}nsicke} BT},
  \bibinfo{author}{{Geier} S}, \bibinfo{author}{{Kupfer} T},
  \bibinfo{author}{{Marsh} TR}, \bibinfo{author}{{Nelemans} G},
  \bibinfo{author}{{O'Toole} SJ}, \bibinfo{author}{{{\O}stensen} RH},
  \bibinfo{author}{{Smalley} B} and  \bibinfo{author}{{West} RG}
  (\bibinfo{year}{2011}), \bibinfo{month}{Dec.}
\bibinfo{title}{{Discovery of a stripped red giant core in a bright eclipsing
  binary system}}.
\bibinfo{journal}{{\em MNRAS}} \bibinfo{volume}{418} (\bibinfo{number}{2}):
  \bibinfo{pages}{1156--1164}.
  \bibinfo{doi}{\doi{10.1111/j.1365-2966.2011.19567.x}}.
\eprint{1107.4986}.

\bibtype{Article}%
\bibitem[{Maxted} et al.(2013)]{2013Natur.498..463M}
\bibinfo{author}{{Maxted} PFL}, \bibinfo{author}{{Serenelli} AM},
  \bibinfo{author}{{Miglio} A}, \bibinfo{author}{{Marsh} TR},
  \bibinfo{author}{{Heber} U}, \bibinfo{author}{{Dhillon} VS},
  \bibinfo{author}{{Littlefair} S}, \bibinfo{author}{{Copperwheat} C},
  \bibinfo{author}{{Smalley} B}, \bibinfo{author}{{Breedt} E} and
  \bibinfo{author}{{Schaffenroth} V} (\bibinfo{year}{2013}),
  \bibinfo{month}{Jun.}
\bibinfo{title}{{Multi-periodic pulsations of a stripped red-giant star in an
  eclipsing binary system}}.
\bibinfo{journal}{{\em Nature}} \bibinfo{volume}{498} (\bibinfo{number}{7455}):
  \bibinfo{pages}{463--465}. \bibinfo{doi}{\doi{10.1038/nature12192}}.
\eprint{1307.1654}.

\bibtype{Article}%
\bibitem[{Mengel} et al.(1976)]{1976ApJ...204..488M}
\bibinfo{author}{{Mengel} JG}, \bibinfo{author}{{Norris} J} and
  \bibinfo{author}{{Gross} PG} (\bibinfo{year}{1976}), \bibinfo{month}{Mar.}
\bibinfo{title}{{Binary Hypothesis for the Subdwarf B Stars}}.
\bibinfo{journal}{{\em ApJ}} \bibinfo{volume}{204}: \bibinfo{pages}{488--492}.
  \bibinfo{doi}{\doi{10.1086/154193}}.

\bibtype{Article}%
\bibitem[{Miller Bertolami} et al.(2022)]{2022MNRAS.511L..60M}
\bibinfo{author}{{Miller Bertolami} MM}, \bibinfo{author}{{Battich} T},
  \bibinfo{author}{{C{\'o}rsico} AH}, \bibinfo{author}{{Althaus} LG} and
  \bibinfo{author}{{Wachlin} FC} (\bibinfo{year}{2022}), \bibinfo{month}{Mar.}
\bibinfo{title}{{An evolutionary channel for CO-rich and pulsating He-rich
  subdwarfs}}.
\bibinfo{journal}{{\em MNRAS}} \bibinfo{volume}{511} (\bibinfo{number}{1}):
  \bibinfo{pages}{L60--L65}. \bibinfo{doi}{\doi{10.1093/mnrasl/slab134}}.
\eprint{2202.05635}.

\bibtype{Article}%
\bibitem[{N{\'e}meth} et al.(2012)]{2012MNRAS.427.2180N}
\bibinfo{author}{{N{\'e}meth} P}, \bibinfo{author}{{Kawka} A} and
  \bibinfo{author}{{Vennes} S} (\bibinfo{year}{2012}), \bibinfo{month}{Dec.}
\bibinfo{title}{{A selection of hot subluminous stars in the GALEX survey - II.
  Subdwarf atmospheric parameters}}.
\bibinfo{journal}{{\em MNRAS}} \bibinfo{volume}{427} (\bibinfo{number}{3}):
  \bibinfo{pages}{2180--2211}.
  \bibinfo{doi}{\doi{10.1111/j.1365-2966.2012.22009.x}}.
\eprint{1211.0323}.

\bibtype{Article}%
\bibitem[{N{\'e}meth} et al.(2021)]{2021A&A...653A...3N}
\bibinfo{author}{{N{\'e}meth} P}, \bibinfo{author}{{Vos} J},
  \bibinfo{author}{{Molina} F} and  \bibinfo{author}{{Bastian} A}
  (\bibinfo{year}{2021}), \bibinfo{month}{Sep.}
\bibinfo{title}{{The first heavy-metal hot subdwarf composite binary SB 744}}.
\bibinfo{journal}{{\em A\&A}} \bibinfo{volume}{653}, \bibinfo{eid}{A3}.
  \bibinfo{doi}{\doi{10.1051/0004-6361/202141442}}.
\eprint{2107.03270}.

\bibtype{Article}%
\bibitem[{Neunteufel} et al.(2022)]{2022A&A...663A..91N}
\bibinfo{author}{{Neunteufel} P}, \bibinfo{author}{{Preece} H},
  \bibinfo{author}{{Kruckow} M}, \bibinfo{author}{{Geier} S},
  \bibinfo{author}{{Hamers} AS}, \bibinfo{author}{{Justham} S} and
  \bibinfo{author}{{Podsiadlowski} P} (\bibinfo{year}{2022}),
  \bibinfo{month}{Jul.}
\bibinfo{title}{{Properties and applications of a predicted population of
  runaway He-sdO/B stars ejected from single degenerate He-donor SNe}}.
\bibinfo{journal}{{\em A\&A}} \bibinfo{volume}{663}, \bibinfo{eid}{A91}.
  \bibinfo{doi}{\doi{10.1051/0004-6361/202142864}}.
\eprint{2112.07469}.

\bibtype{Article}%
\bibitem[{Pakmor} et al.(2024)]{2024arXiv240702566P}
\bibinfo{author}{{Pakmor} R}, \bibinfo{author}{{Pelisoli} I},
  \bibinfo{author}{{Justham} S}, \bibinfo{author}{{Rajamuthukumar} AS},
  \bibinfo{author}{{R{\"o}pke} FK}, \bibinfo{author}{{Schneider} FRN},
  \bibinfo{author}{{de Mink} SE}, \bibinfo{author}{{Ohlmann} ST},
  \bibinfo{author}{{Podsiadlowski} P}, \bibinfo{author}{{Moran Fraile} J},
  \bibinfo{author}{{Vetter} M} and  \bibinfo{author}{{Andrassy} R}
  (\bibinfo{year}{2024}), \bibinfo{month}{Jul.}
\bibinfo{title}{{Large-scale ordered magnetic fields generated in mergers of
  helium white dwarfs}}.
\bibinfo{journal}{{\em arXiv e-prints}} ,
  \bibinfo{eid}{arXiv:2407.02566}\bibinfo{doi}{\doi{10.48550/arXiv.2407.02566}}.
\eprint{2407.02566}.

\bibtype{Article}%
\bibitem[{Pelisoli} et al.(2020)]{2020A&A...642A.180P}
\bibinfo{author}{{Pelisoli} I}, \bibinfo{author}{{Vos} J},
  \bibinfo{author}{{Geier} S}, \bibinfo{author}{{Schaffenroth} V} and
  \bibinfo{author}{{Baran} AS} (\bibinfo{year}{2020}), \bibinfo{month}{Oct.}
\bibinfo{title}{{Alone but not lonely: Observational evidence that binary
  interaction is always required to form hot subdwarf stars}}.
\bibinfo{journal}{{\em A\&A}} \bibinfo{volume}{642}, \bibinfo{eid}{A180}.
  \bibinfo{doi}{\doi{10.1051/0004-6361/202038473}}.
\eprint{2008.07522}.

\bibtype{Article}%
\bibitem[{Pelisoli} et al.(2022)]{2022MNRAS.515.2496P}
\bibinfo{author}{{Pelisoli} I}, \bibinfo{author}{{Dorsch} M},
  \bibinfo{author}{{Heber} U}, \bibinfo{author}{{G{\"a}nsicke} B},
  \bibinfo{author}{{Geier} S}, \bibinfo{author}{{Kupfer} T},
  \bibinfo{author}{{N{\'e}meth} P}, \bibinfo{author}{{Scaringi} S} and
  \bibinfo{author}{{Schaffenroth} V} (\bibinfo{year}{2022}),
  \bibinfo{month}{Sep.}
\bibinfo{title}{{Discovery and analysis of three magnetic hot subdwarf stars:
  evidence for merger-induced magnetic fields}}.
\bibinfo{journal}{{\em MNRAS}} \bibinfo{volume}{515} (\bibinfo{number}{2}):
  \bibinfo{pages}{2496--2510}. \bibinfo{doi}{\doi{10.1093/mnras/stac1069}}.
\eprint{2204.06575}.

\bibtype{Article}%
\bibitem[{Peng} et al.(2024)]{2024NewA..10702153P}
\bibinfo{author}{{Peng} Y}, \bibinfo{author}{{Wang} K} and
  \bibinfo{author}{{Ren} A} (\bibinfo{year}{2024}), \bibinfo{month}{Apr.}
\bibinfo{title}{{Discovery of four new EL CVn-type binaries in the Gaia
  eclipsing binaries}}.
\bibinfo{journal}{{\em NA}} \bibinfo{volume}{107}, \bibinfo{eid}{102153}.
  \bibinfo{doi}{\doi{10.1016/j.newast.2023.102153}}.
\eprint{2310.19307}.

\bibtype{Article}%
\bibitem[{Piersanti} et al.(2024)]{2024arXiv240517896P}
\bibinfo{author}{{Piersanti} L}, \bibinfo{author}{{Yungelson} LR} and
  \bibinfo{author}{{Bravo} E} (\bibinfo{year}{2024}), \bibinfo{month}{May}.
\bibinfo{title}{{The expected evolution of the binary system PTF
  J2238+743015.1}}.
\bibinfo{journal}{{\em arXiv e-prints}} ,
  \bibinfo{eid}{arXiv:2405.17896}\bibinfo{doi}{\doi{10.48550/arXiv.2405.17896}}.
\eprint{2405.17896}.

\bibtype{Article}%
\bibitem[{Pietrukowicz} et al.(2024)]{2024arXiv240416089P}
\bibinfo{author}{{Pietrukowicz} P}, \bibinfo{author}{{Latour} M},
  \bibinfo{author}{{Soszynski} I}, \bibinfo{author}{{Di Mille} F},
  \bibinfo{author}{{King} PS}, \bibinfo{author}{{Angeloni} R},
  \bibinfo{author}{{Poleski} R}, \bibinfo{author}{{Udalski} A},
  \bibinfo{author}{{Szymanski} MK}, \bibinfo{author}{{Ulaczyk} K},
  \bibinfo{author}{{Kozlowski} S}, \bibinfo{author}{{Skowron} J},
  \bibinfo{author}{{Skowron} DM}, \bibinfo{author}{{Mroz} P},
  \bibinfo{author}{{Rybicki} K}, \bibinfo{author}{{Iwanek} P},
  \bibinfo{author}{{Wrona} M} and  \bibinfo{author}{{Gromadzki} M}
  (\bibinfo{year}{2024}), \bibinfo{month}{Apr.}
\bibinfo{title}{{Observational parameters of Blue Large-Amplitude Pulsators}}.
\bibinfo{journal}{{\em arXiv e-prints}} ,
  \bibinfo{eid}{arXiv:2404.16089}\bibinfo{doi}{\doi{10.48550/arXiv.2404.16089}}.
\eprint{2404.16089}.

\bibtype{Inproceedings}%
\bibitem[{Podsiadlowski} et al.(2008)]{2008ASPC..392...15P}
\bibinfo{author}{{Podsiadlowski} P}, \bibinfo{author}{{Han} Z},
  \bibinfo{author}{{Lynas-Gray} AE} and  \bibinfo{author}{{Brown} D}
  (\bibinfo{year}{2008}), \bibinfo{month}{Jan.}, \bibinfo{title}{{Hot Subdwarfs
  in Binaries as the Source of the Far-UV Excess in Elliptical Galaxies}},
  \bibinfo{editor}{{Heber} U}, \bibinfo{editor}{{Jeffery} CS} and
  \bibinfo{editor}{{Napiwotzki} R}, (Eds.), \bibinfo{booktitle}{Hot Subdwarf
  Stars and Related Objects}, \bibinfo{series}{Astronomical Society of the
  Pacific Conference Series}, \bibinfo{volume}{392}, pp.~\bibinfo{pages}{15},
  \eprint{0808.0574}.

\bibtype{Article}%
\bibitem[{Preece} et al.(2019)]{2019MNRAS.485.2889P}
\bibinfo{author}{{Preece} HP}, \bibinfo{author}{{Tout} CA} and
  \bibinfo{author}{{Jeffery} CS} (\bibinfo{year}{2019}), \bibinfo{month}{May}.
\bibinfo{title}{{Convection physics and tidal synchronization of the subdwarf
  binary NY Virginis}}.
\bibinfo{journal}{{\em MNRAS}} \bibinfo{volume}{485} (\bibinfo{number}{2}):
  \bibinfo{pages}{2889--2894}. \bibinfo{doi}{\doi{10.1093/mnras/stz547}}.
\eprint{1903.06176}.

\bibtype{Article}%
\bibitem[{Preece} et al.(2022)]{2022MNRAS.517.2111P}
\bibinfo{author}{{Preece} HP}, \bibinfo{author}{{Hamers} AS},
  \bibinfo{author}{{Battich} T} and  \bibinfo{author}{{Rajamuthukumar} AS}
  (\bibinfo{year}{2022}), \bibinfo{month}{Dec.}
\bibinfo{title}{{Forming hot subluminous stars from hierarchical triples - I.
  The role of an outer tertiary on formation channels}}.
\bibinfo{journal}{{\em MNRAS}} \bibinfo{volume}{517} (\bibinfo{number}{2}):
  \bibinfo{pages}{2111--2120}. \bibinfo{doi}{\doi{10.1093/mnras/stac2798}}.
\eprint{2209.11327}.

\bibtype{Inproceedings}%
\bibitem[{Przybilla} et al.(2011)]{2011JPhCS.328a2015P}
\bibinfo{author}{{Przybilla} N}, \bibinfo{author}{{Nieva} MF} and
  \bibinfo{author}{{Butler} K} (\bibinfo{year}{2011}), \bibinfo{month}{Dec.},
  \bibinfo{title}{{Testing common classical LTE and NLTE model atmosphere and
  line-formation codes for quantitative spectroscopy of early-type stars}},
  \bibinfo{booktitle}{Journal of Physics Conference Series},
  \bibinfo{series}{Journal of Physics Conference Series},
  \bibinfo{volume}{328}, \bibinfo{publisher}{IOP}, pp. \bibinfo{pages}{012015},
  \eprint{1111.1445}.

\bibtype{Article}%
\bibitem[{Pulley} et al.(2022)]{2022MNRAS.514.5725P}
\bibinfo{author}{{Pulley} D}, \bibinfo{author}{{Sharp} ID},
  \bibinfo{author}{{Mallett} J} and  \bibinfo{author}{{von Harrach} S}
  (\bibinfo{year}{2022}), \bibinfo{month}{Aug.}
\bibinfo{title}{{Eclipse timing variations in post-common envelope binaries:
  Are they a reliable indicator of circumbinary companions?}}
\bibinfo{journal}{{\em MNRAS}} \bibinfo{volume}{514} (\bibinfo{number}{4}):
  \bibinfo{pages}{5725--5738}. \bibinfo{doi}{\doi{10.1093/mnras/stac1676}}.
\eprint{2206.06919}.

\bibtype{Article}%
\bibitem[{Rebassa-Mansergas} et al.(2024)]{2024A&A...686A.221R}
\bibinfo{author}{{Rebassa-Mansergas} A}, \bibinfo{author}{{Hollands} M},
  \bibinfo{author}{{Parsons} SG}, \bibinfo{author}{{Althaus} LG},
  \bibinfo{author}{{Pelisoli} I}, \bibinfo{author}{{Irawati} P},
  \bibinfo{author}{{Raddi} R}, \bibinfo{author}{{Camisassa} ME} and
  \bibinfo{author}{{Torres} S} (\bibinfo{year}{2024}), \bibinfo{month}{Jun.}
\bibinfo{title}{{J0526+5934: A peculiar ultra-short-period double white
  dwarf}}.
\bibinfo{journal}{{\em A\&A}} \bibinfo{volume}{686}, \bibinfo{eid}{A221}.
  \bibinfo{doi}{\doi{10.1051/0004-6361/202449519}}.
\eprint{2402.04443}.

\bibtype{Article}%
\bibitem[{Reed} et al.(2021)]{2021MNRAS.507.4178R}
\bibinfo{author}{{Reed} MD}, \bibinfo{author}{{Slayton} A},
  \bibinfo{author}{{Baran} AS}, \bibinfo{author}{{Telting} JH},
  \bibinfo{author}{{{\O}stensen} RH}, \bibinfo{author}{{Jeffery} CS},
  \bibinfo{author}{{Uzundag} M} and  \bibinfo{author}{{Sanjayan} S}
  (\bibinfo{year}{2021}), \bibinfo{month}{Nov.}
\bibinfo{title}{{Pulsating subdwarf B stars observed with K2 during Campaign 7
  and an examination of seismic group properties}}.
\bibinfo{journal}{{\em MNRAS}} \bibinfo{volume}{507} (\bibinfo{number}{3}):
  \bibinfo{pages}{4178--4195}. \bibinfo{doi}{\doi{10.1093/mnras/stab2405}}.
\eprint{2108.07145}.

\bibtype{Article}%
\bibitem[{Saffer} et al.(1998)]{1998ApJ...502..394S}
\bibinfo{author}{{Saffer} RA}, \bibinfo{author}{{Livio} M} and
  \bibinfo{author}{{Yungelson} LR} (\bibinfo{year}{1998}),
  \bibinfo{month}{Jul.}
\bibinfo{title}{{Close Binary White Dwarf Systems: Numerous New Detections and
  Their Interpretation}}.
\bibinfo{journal}{{\em ApJ}} \bibinfo{volume}{502} (\bibinfo{number}{1}):
  \bibinfo{pages}{394--407}. \bibinfo{doi}{\doi{10.1086/305907}}.
\eprint{astro-ph/9802356}.

\bibtype{Article}%
\bibitem[{Schaffenroth} et al.(2022)]{2022A&A...666A.182S}
\bibinfo{author}{{Schaffenroth} V}, \bibinfo{author}{{Pelisoli} I},
  \bibinfo{author}{{Barlow} BN}, \bibinfo{author}{{Geier} S} and
  \bibinfo{author}{{Kupfer} T} (\bibinfo{year}{2022}), \bibinfo{month}{Oct.}
\bibinfo{title}{{Hot subdwarfs in close binaries observed from space. I.
  Orbital, atmospheric, and absolute parameters, and the nature of their
  companions}}.
\bibinfo{journal}{{\em A\&A}} \bibinfo{volume}{666}, \bibinfo{eid}{A182}.
  \bibinfo{doi}{\doi{10.1051/0004-6361/202244214}}.
\eprint{2207.02001}.

\bibtype{Article}%
\bibitem[{Schaffenroth} et al.(2023)]{2023A&A...673A..90S}
\bibinfo{author}{{Schaffenroth} V}, \bibinfo{author}{{Barlow} BN},
  \bibinfo{author}{{Pelisoli} I}, \bibinfo{author}{{Geier} S} and
  \bibinfo{author}{{Kupfer} T} (\bibinfo{year}{2023}), \bibinfo{month}{May}.
\bibinfo{title}{{Hot subdwarfs in close binaries observed from space. II.
  Analyses of the light variations}}.
\bibinfo{journal}{{\em A\&A}} \bibinfo{volume}{673}, \bibinfo{eid}{A90}.
  \bibinfo{doi}{\doi{10.1051/0004-6361/202244697}}.
\eprint{2302.12507}.

\bibtype{Article}%
\bibitem[{Silvotti} et al.(2022)]{2022MNRAS.511.2201S}
\bibinfo{author}{{Silvotti} R}, \bibinfo{author}{{N{\'e}meth} P},
  \bibinfo{author}{{Telting} JH}, \bibinfo{author}{{Baran} AS},
  \bibinfo{author}{{{\O}stensen} RH}, \bibinfo{author}{{Ostrowski} J},
  \bibinfo{author}{{Sahoo} SK} and  \bibinfo{author}{{Prins} S}
  (\bibinfo{year}{2022}), \bibinfo{month}{Apr.}
\bibinfo{title}{{Filling the gap between synchronized and non-synchronized sdBs
  in short-period sdBV+dM binaries with TESS: TIC 137608661, a new system with
  a well-defined rotational splitting}}.
\bibinfo{journal}{{\em MNRAS}} \bibinfo{volume}{511} (\bibinfo{number}{2}):
  \bibinfo{pages}{2201--2217}. \bibinfo{doi}{\doi{10.1093/mnras/stac160}}.
\eprint{2201.06559}.

\bibtype{Article}%
\bibitem[{Su} et al.(2024)]{2024arXiv240717887S}
\bibinfo{author}{{Su} W}, \bibinfo{author}{{Charpinet} S},
  \bibinfo{author}{{Latour} M}, \bibinfo{author}{{Zong} W},
  \bibinfo{author}{{Green} EM} and  \bibinfo{author}{{Li} G}
  (\bibinfo{year}{2024}), \bibinfo{month}{Jul.}
\bibinfo{title}{{TIC441725813: A new bright hybrid sdB pulsator with
  differential core/envelope rotation}}.
\bibinfo{journal}{{\em arXiv e-prints}} ,
  \bibinfo{eid}{arXiv:2407.17887}\bibinfo{doi}{\doi{10.48550/arXiv.2407.17887}}.
\eprint{2407.17887}.

\bibtype{Article}%
\bibitem[{Van Grootel} et al.(2013)]{2013A&A...553A..97V}
\bibinfo{author}{{Van Grootel} V}, \bibinfo{author}{{Charpinet} S},
  \bibinfo{author}{{Brassard} P}, \bibinfo{author}{{Fontaine} G} and
  \bibinfo{author}{{Green} EM} (\bibinfo{year}{2013}), \bibinfo{month}{May}.
\bibinfo{title}{{Third generation stellar models for asteroseismology of hot B
  subdwarf stars. A test of accuracy with the pulsating eclipsing binary PG
  1336-018}}.
\bibinfo{journal}{{\em A\&A}} \bibinfo{volume}{553}, \bibinfo{eid}{A97}.
  \bibinfo{doi}{\doi{10.1051/0004-6361/201220896}}.
\eprint{1303.3368}.

\bibtype{Article}%
\bibitem[{van Kerkwijk} et al.(1996)]{1996ApJ...467L..89V}
\bibinfo{author}{{van Kerkwijk} MH}, \bibinfo{author}{{Bergeron} P} and
  \bibinfo{author}{{Kulkarni} SR} (\bibinfo{year}{1996}), \bibinfo{month}{Aug.}
\bibinfo{title}{{The Masses of the Millisecond Pulsar J1012+5307 and Its White
  Dwarf Companion}}.
\bibinfo{journal}{{\em ApJL}} \bibinfo{volume}{467}: \bibinfo{pages}{L89}.
  \bibinfo{doi}{\doi{10.1086/310209}}.
\eprint{astro-ph/9606045}.

\bibtype{Article}%
\bibitem[{Vos} et al.(2015)]{2015A&A...579A..49V}
\bibinfo{author}{{Vos} J}, \bibinfo{author}{{{\O}stensen} RH},
  \bibinfo{author}{{Marchant} P} and  \bibinfo{author}{{Van Winckel} H}
  (\bibinfo{year}{2015}), \bibinfo{month}{Jul.}
\bibinfo{title}{{Testing eccentricity pumping mechanisms to model eccentric
  long-period sdB binaries with MESA}}.
\bibinfo{journal}{{\em A\&A}} \bibinfo{volume}{579}, \bibinfo{eid}{A49}.
  \bibinfo{doi}{\doi{10.1051/0004-6361/201526019}}.
\eprint{1505.03293}.

\bibtype{Article}%
\bibitem[{Vos} et al.(2018)]{2018MNRAS.473..693V}
\bibinfo{author}{{Vos} J}, \bibinfo{author}{{N{\'e}meth} P},
  \bibinfo{author}{{Vu{\v{c}}kovi{\'c}} M}, \bibinfo{author}{{{\O}stensen} R}
  and  \bibinfo{author}{{Parsons} S} (\bibinfo{year}{2018}),
  \bibinfo{month}{Jan.}
\bibinfo{title}{{Composite hot subdwarf binaries - I. The spectroscopically
  confirmed sdB sample}}.
\bibinfo{journal}{{\em MNRAS}} \bibinfo{volume}{473} (\bibinfo{number}{1}):
  \bibinfo{pages}{693--709}. \bibinfo{doi}{\doi{10.1093/mnras/stx2198}}.
\eprint{1708.07340}.

\bibtype{Article}%
\bibitem[{Vos} et al.(2020)]{2020A&A...641A.163V}
\bibinfo{author}{{Vos} J}, \bibinfo{author}{{Bobrick} A} and
  \bibinfo{author}{{Vu{\v{c}}kovi{\'c}} M} (\bibinfo{year}{2020}),
  \bibinfo{month}{Sep.}
\bibinfo{title}{{Observed binary populations reflect the Galactic history.
  Explaining the orbital period-mass ratio relation in wide hot subdwarf
  binaries}}.
\bibinfo{journal}{{\em A\&A}} \bibinfo{volume}{641}, \bibinfo{eid}{A163}.
  \bibinfo{doi}{\doi{10.1051/0004-6361/201937195}}.
\eprint{2003.05665}.

\bibtype{Article}%
\bibitem[{Vos} et al.(2021)]{2021A&A...655A..43V}
\bibinfo{author}{{Vos} J}, \bibinfo{author}{{Pelisoli} I},
  \bibinfo{author}{{Budaj} J}, \bibinfo{author}{{Reindl} N},
  \bibinfo{author}{{Schaffenroth} V}, \bibinfo{author}{{Bobrick} A},
  \bibinfo{author}{{Geier} S}, \bibinfo{author}{{Hermes} J},
  \bibinfo{author}{{Nemeth} P}, \bibinfo{author}{{{\O}stensen} R},
  \bibinfo{author}{{Reding} JS}, \bibinfo{author}{{Uzundag} M} and
  \bibinfo{author}{{Vu{\v{c}}kovi{\'c}} M} (\bibinfo{year}{2021}),
  \bibinfo{month}{Nov.}
\bibinfo{title}{{Looking into the cradle of the grave: J22564-5910, a potential
  young post-merger hot subdwarf}}.
\bibinfo{journal}{{\em A\&A}} \bibinfo{volume}{655}, \bibinfo{eid}{A43}.
  \bibinfo{doi}{\doi{10.1051/0004-6361/202140391}}.
\eprint{2106.03363}.

\bibtype{Article}%
\bibitem[{Vu{\v{c}}kovi{\'c}} et al.(2007)]{2007A&A...471..605V}
\bibinfo{author}{{Vu{\v{c}}kovi{\'c}} M}, \bibinfo{author}{{Aerts} C},
  \bibinfo{author}{{{\"O}stensen} R}, \bibinfo{author}{{Nelemans} G},
  \bibinfo{author}{{Hu} H}, \bibinfo{author}{{Jeffery} CS},
  \bibinfo{author}{{Dhillon} VS} and  \bibinfo{author}{{Marsh} TR}
  (\bibinfo{year}{2007}), \bibinfo{month}{Aug.}
\bibinfo{title}{{The binary properties of the pulsating subdwarf B eclipsing
  binary <ASTROBJ>PG 1336-018</ASTROBJ> (NY Virginis)}}.
\bibinfo{journal}{{\em A\&A}} \bibinfo{volume}{471} (\bibinfo{number}{2}):
  \bibinfo{pages}{605--615}. \bibinfo{doi}{\doi{10.1051/0004-6361:20077179}}.
\eprint{0706.3363}.

\bibtype{Article}%
\bibitem[{Webbink}(1984)]{1984ApJ...277..355W}
\bibinfo{author}{{Webbink} RF} (\bibinfo{year}{1984}), \bibinfo{month}{Feb.}
\bibinfo{title}{{Double white dwarfs as progenitors of R Coronae Borealis stars
  and type I supernovae.}}
\bibinfo{journal}{{\em ApJ}} \bibinfo{volume}{277}: \bibinfo{pages}{355--360}.
  \bibinfo{doi}{\doi{10.1086/161701}}.

\bibtype{Article}%
\bibitem[{Werner} et al.(2022)]{2022MNRAS.511L..66W}
\bibinfo{author}{{Werner} K}, \bibinfo{author}{{Reindl} N},
  \bibinfo{author}{{Geier} S} and  \bibinfo{author}{{Pritzkuleit} M}
  (\bibinfo{year}{2022}), \bibinfo{month}{Mar.}
\bibinfo{title}{{Discovery of hot subdwarfs covered with helium-burning ash}}.
\bibinfo{journal}{{\em MNRAS}} \bibinfo{volume}{511} (\bibinfo{number}{1}):
  \bibinfo{pages}{L66--L71}. \bibinfo{doi}{\doi{10.1093/mnrasl/slac005}}.
\eprint{2202.05633}.

\bibtype{Article}%
\bibitem[{Wu} et al.(2018)]{2018A&A...618A..14W}
\bibinfo{author}{{Wu} Y}, \bibinfo{author}{{Chen} X}, \bibinfo{author}{{Li} Z}
  and  \bibinfo{author}{{Han} Z} (\bibinfo{year}{2018}), \bibinfo{month}{Oct.}
\bibinfo{title}{{Formation of hot subdwarf B stars with neutron star
  components}}.
\bibinfo{journal}{{\em A\&A}} \bibinfo{volume}{618}, \bibinfo{eid}{A14}.
  \bibinfo{doi}{\doi{10.1051/0004-6361/201832686}}.
\eprint{1808.03402}.

\bibtype{Article}%
\bibitem[{Yu} et al.(2021)]{2021MNRAS.504.2670Y}
\bibinfo{author}{{Yu} J}, \bibinfo{author}{{Zhang} X} and
  \bibinfo{author}{{L{\"u}} G} (\bibinfo{year}{2021}), \bibinfo{month}{Jun.}
\bibinfo{title}{{Post-merger evolution of double helium white dwarfs and
  distribution of helium-rich hot subdwarfs}}.
\bibinfo{journal}{{\em MNRAS}} \bibinfo{volume}{504} (\bibinfo{number}{2}):
  \bibinfo{pages}{2670--2674}. \bibinfo{doi}{\doi{10.1093/mnras/stab1063}}.
\eprint{2104.06817}.

\bibtype{Article}%
\bibitem[{Zhang} and {Jeffery}(2012)]{2012MNRAS.419..452Z}
\bibinfo{author}{{Zhang} X} and  \bibinfo{author}{{Jeffery} CS}
  (\bibinfo{year}{2012}), \bibinfo{month}{Jan.}
\bibinfo{title}{{Evolutionary models for double helium white dwarf mergers and
  the formation of helium-rich hot subdwarfs}}.
\bibinfo{journal}{{\em MNRAS}} \bibinfo{volume}{419} (\bibinfo{number}{1}):
  \bibinfo{pages}{452--464}.
  \bibinfo{doi}{\doi{10.1111/j.1365-2966.2011.19711.x}}.

\bibtype{Article}%
\bibitem[{Zhang} et al.(2023)]{2023ApJ...959...24Z}
\bibinfo{author}{{Zhang} X}, \bibinfo{author}{{Jeffery} CS},
  \bibinfo{author}{{Su} J} and  \bibinfo{author}{{Bi} S}
  (\bibinfo{year}{2023}), \bibinfo{month}{Dec.}
\bibinfo{title}{{The Formation of Blue Large-amplitude Pulsators from
  White-dwarf Main-sequence Star Mergers}}.
\bibinfo{journal}{{\em ApJ}} \bibinfo{volume}{959} (\bibinfo{number}{1}),
  \bibinfo{eid}{24}. \bibinfo{doi}{\doi{10.3847/1538-4357/ad0a65}}.
\eprint{2311.07812}.

\end{thebibliography*}

\end{document}